\documentclass{article}

\usepackage{PRIMEarxiv}

\usepackage[utf8]{inputenc} 
\usepackage[T1]{fontenc}    
\usepackage{hyperref}       
\usepackage{url}            
\usepackage{booktabs}       
\usepackage{amsfonts}       
\usepackage{nicefrac}       
\usepackage{microtype}      
\usepackage{lipsum}
\usepackage{fancyhdr}       
\usepackage{graphicx}       

\usepackage{tabularx}
\usepackage{multirow}

\usepackage{amsmath,amssymb}
\usepackage{amsthm}      

\usepackage{enumitem}    

\usepackage{mathtools}

\title{Why the Brain Consolidates:\\Predictive Forgetting for Optimal Generalisation}

\author{Zafeirios Fountas$^{1,}$\thanks{Corresponding author: zafeirios.fountas@huawei.com}~, Adnan Oomerjee$^{1,2}$, Haitham Bou-Ammar$^{1,2}$, Jun Wang$^{2}$, Neil Burgess$^{3,4}$ \\
1. Huawei Noah’s Ark Lab, London, UK\\
2. AI Centre, Department of Computer Science, University College London, London, UK\\
3. UCL Institute of Cognitive Neuroscience, University College London, London, UK\\
4. UCL Queen Square Institute of Neurology, University College London, London, UK}

\begin{document}

\maketitle

\begin{abstract}

Standard accounts of memory consolidation emphasise the stabilisation of stored representations, but struggle to explain representational drift, semanticisation, or the necessity of offline replay. Here we propose that high-capacity neocortical networks optimise stored representations for generalisation by reducing complexity via \emph{predictive forgetting}, i.e. the selective retention of experienced information that predicts future outcomes or experience. We show that predictive forgetting formally improves information-theoretic generalisation bounds on stored representations. Under high-fidelity encoding constraints, such compression is generally unattainable in a single pass; high-capacity networks therefore benefit from temporally separated, iterative refinement of stored traces without re-accessing sensory input. We demonstrate this capacity dependence with simulations in autoencoder-based neocortical models, biologically plausible predictive coding circuits, and Transformer-based language models, and derive quantitative predictions for consolidation-dependent changes in neural representational geometry. These results identify a computational role for off-line consolidation beyond stabilisation, showing that outcome-conditioned compression optimises the retention-generalisation trade-off.

\end{abstract}

\section{Introduction}

A fundamental aspect of intelligence is the capacity to leverage previous experience to solve novel problems. This ability to generalise determines how effectively learned knowledge transfers beyond the specific contexts in which it was acquired. Without it, organisms struggle to navigate changing environments whilst artificial systems become brittle and data-inefficient~\cite{mcclelland1995, lake2017}. In machine learning, the tension between learning and retention manifests dramatically as \emph{catastrophic forgetting}: sequential training on multiple tasks can cause near-complete erasure of earlier learning~\cite{mccloskey1989, ratcliff1990, french1999}. Yet perfect retention of every detail is itself maladaptive. In humans, near-perfect autobiographical memory can impair function by overwhelming cognitive resources with irrelevant details~\cite{parker2006, leport2016}, suggesting that active forgetting is intrinsic to adaptive memory function~\cite{richards2017persistence, hardt2013decay, davis2017biology}. Normative accounts formalise this link, showing that complexity reduction improves generalisation across memory~\cite{sun2023organizing} and control~\cite{moskovitz2024understanding} domains. Critically, this compression targets the neocortical generative model that supports generalisation, while veridical episodic details may persist in hippocampus~\cite{spens2024generative,nicholas2026episodic}. This challenge extends to artificial intelligence, where large language models (LLMs) face severe computational constraints as context windows grow~\cite{vaswani2017, liu2024lost}. These considerations reveal a fundamental fidelity-generalisation conflict in learning: the tension between retaining high-fidelity details for immediate demands and enforcing the compression necessary for robust generalisation.

The transformation from detailed episodic memory to compressed semantic knowledge is commonly attributed to \emph{consolidation}, an offline reorganisation of memory traces that operates independently of new sensory input~\cite{dudai2004,squire2015}. Classical theories, originating with Marr~\cite{marr1971simple}, describe 
a gradual shift from hippocampal to neocortical 
dependence~\cite{squire1995, mcclelland1995}, often framed as schema formation~\cite{lewis2011}, sleep-dependent triage~\cite{stickgold2013, payne2008sleep}, or synaptic renormalisation~\cite{tononi2014sleep,diering2017homer1a}. Recent computational models have demonstrated how generative networks trained via hippocampal replay can implement this reorganisation to explain memory construction and schema-based distortions~\cite{spens2024generative}. Normative accounts have begun to formalise the computational logic: Nagy et al.~\cite{nagy2025adaptive} applied rate-distortion theory to frame memory as adaptive compression; Sun et al.~\cite{sun2023organizing} showed that consolidation should be regulated by outcome unpredictability; Moskovitz et al.~\cite{moskovitz2024understanding} demonstrated that complexity reduction governs the trade-off between deliberative and habitual behaviour; and Spens et al.~\cite{spens2025modelling} proposed that meta-controllers can learn adaptive curricula to orchestrate these processes. Yet while these frameworks explain \textit{when} to consolidate or \textit{how} to schedule it, they do not explain \textit{why} the memory trace itself must be actively transformed, including the fundamental questions of what representational objective consolidation should optimise and why achieving it requires temporal separation from encoding.

Here, we propose that consolidation into neocortex optimises generalisation by implementing what we term \emph{predictive forgetting}, the progressive elimination of information about sensory inputs that does not predict future outcomes or experience. In essence, this mechanism selectively discards the incidental details of an experience ($X$) that do not predict its consequences ($Y$)---details that may be retained by hippocampal episodic memory~\cite{spens2024generative, nicholas2026episodic}. Drawing on recent advances in learning theory~\cite{kawaguchi2023}, we show that minimising this outcome-conditional mutual information between sensory inputs and stored memory traces provably tightens generalisation bounds.

We demonstrate that the pressure for this compression scales with \emph{representational capacity}. In low-capacity regimes, architectural bottlenecks naturally filter noise, avoiding the need for repeated replay. However, in high-capacity regimes, characteristic of mammalian neocortex, the system possesses sufficient degrees of freedom to memorise sensory noise, and single-pass learning cannot generally achieve the required compression. Iterative offline replay (the reactivation of neural activity patterns during rest or sleep~\cite{wilson1994reactivation, girardeau2009selective}) can force downstream readouts to learn from compressed codes, suppressing overfitting. This provides a normative rationale for why high-capacity systems require temporally separated refinement to generalise well. We validate this principle across diverse substrates, from neocortical predictive coding loops to cache refinement in large language models, unifying systems consolidation, representational drift, memory semanticisation, and continual learning stability under a single optimisation objective. We further derive quantitative predictions for consolidation-dependent changes in neural representational geometry testable through longitudinal neuroimaging and sleep manipulations.

\section{Theoretical Framework}

\begin{figure}[t]
    \centering
    \includegraphics[width=0.99\linewidth]{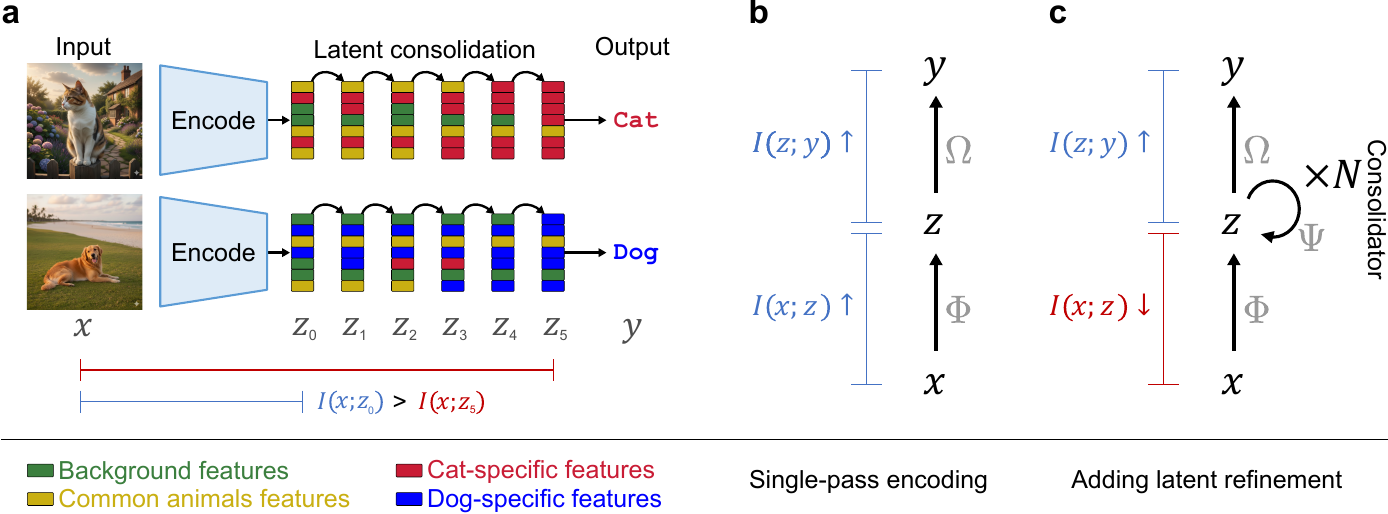}
    \caption{\textbf{Consolidation as predictive forgetting.} 
    \textbf{a,} Conceptual illustration using visual classification. Initial encoding (distributions $Z_0$; lowercase variables denote specific samples) preserves both diagnostic features (cat-specific: red; dog-specific: blue; shared: yellow) and task-irrelevant details (background, lighting: green) to minimise sensory prediction error. Iterative consolidation ($Z_1 \to Z_5$) progressively discards noise (shared and task-irrelevant features) while sharpening category-predictive features.
    \textbf{b,} Online Learning (Wake) acts as standard feedforward learning. The encoder module (parameters $\Phi$) maps inputs $X$ to high-fidelity representations $Z$ to predict targets $Y$. Training objectives that maximise task performance can drive both mutual information terms $I(X;Z)$ and $I(Z;Y)$ upwards; however, a high-capacity readout (parameters $\Omega$) minimises error by memorising the input-specific noise retained in $Z$, leading to overfitting.
    \textbf{c,} Consolidation through iterative latent refinement (modelled as an offline ``sleep'' phase, cf.\ the wake-sleep algorithm~\cite{hinton1995wake}).
    Augmenting the encoder with an offline consolidator $\Psi$ (applied $N$ times) enables a different optimisation: $I(X;Z)$ decreases as $I(Z;Y)$ is maintained (or increases). By actively reducing input dependence, this process enforces an information bottleneck on the downstream readout $\Omega$ (formally constraining the Markov chain $S \to Z \to \Omega$), physically preventing the overfitting that plagues high-capacity single-pass systems and thus tightening the generalisation bound (Equation~\ref{eq:ib-bound}).
    Note that this schematic depicts the consolidation of the neocortical representation used for generalisation; the hippocampal system may additionally retain veridical episodic details not captured by the compressed code.}
    \label{fig:intro}
\end{figure}

Consider a perceptual system learning to distinguish dogs from cats (Fig.~\ref{fig:intro}). During initial encoding, the system captures every detail: fur texture, lighting, background objects. While such rich representation supports immediate recognition, it is inefficient for generalisation, as it includes features irrelevant to the core decision. Ideally, the system should favour diagnostic features (ear shape, facial structure) above incidental details~\cite{tandoc2024object}. However, during rapid online encoding, the system often cannot yet distinguish signal from noise. It must capture more than is ultimately needed, but can then gradually compress its representations offline. We propose that this is a role for consolidation. The neocortical generative model is iteratively refined to emphasise predictive structure, leaving the hippocampus with important veridical details not captured by this compression~\cite{spens2024generative,nicholas2026episodic}. Recent advances in learning theory allow us to formalise this intuition mathematically.

\subsection{The Generalisation Bound}
\label{sec:gen-bound}

Generalisation error is controlled by the trade-off between how well a model fits the training data and how complex its representations are. Recent information-theoretic work has formalised this by showing that the generalisation gap $\Delta$ (the difference between a model's performance on training data and on novel data) is upper-bounded by the amount of input information retained in the memory trace that is not explained by the outcome~\cite{kawaguchi2023}. For a representation $Z$ of input $X$ and predictive target $Y$ (which may be a task label, a required output, or the next input in a sequence), the generalisation gap satisfies:
\begin{equation}
\Delta\;\le\;\tilde{\mathcal{O}}\!\left(\sqrt{\frac{I(X;Z\mid Y)\;+\;C}{n}}\right),
\label{eq:ib-bound}
\end{equation}
where $n$ is the number of training examples in a dataset $S$, and $C$ represents the complexity of the learned encoder parameters $\Phi$. Formally, $C$ is the mutual information between these parameters and the dataset, $I(\Phi;S)$ (see Methods). The critical term is the \textit{conditional mutual information} $I(X;Z\mid Y)$, which measures the information in the memory $Z$ about the input $X$ that remains \textit{after} the outcome $Y$ is known. Reducing this term directly tightens the generalisation bound. In cognitive terms, this corresponds to the removal of episodic idiosyncrasies (e.g., background lighting) from the consolidated neocortical representation, retaining only the semantic core that generalises across contexts.

\subsection{Consolidation as Predictive Forgetting}
\label{sec:consolidation-def}

We formalise neocortical consolidation as \textit{conditional information compression}. Intuitively, this means discarding information about \textit{how} something was experienced while retaining information about \textit{what it predicts}. Consider remembering a restaurant visit: the precise lighting and background noise are observational details ($X$) stored in the initial episodic trace. However, these features are irrelevant to the core prediction of meal quality ($Y$). We propose that consolidation progressively strips away these incidental details. Formally, a mapping $\mathcal{C}:Z\mapsto Z'$ is a consolidation step if it selectively reduces the mutual information with the input conditional on the outcome:
\begin{equation}
    I(X;Z'\mid Y)\;\le\;I(X;Z\mid Y)
    \quad\text{and}\quad
    I(Y;Z')\;\ge\;I(Y;Z)-\varepsilon.
    \label{eq:consolidation}
\end{equation}
The first inequality implements \textit{predictive forgetting}: the elimination of statistical dependence between memories and inputs that does not aid prediction. This objective effectively drives the representation toward a \textit{minimal sufficient statistic for predicting $Y$}--the most compressed form that retains full predictive power about outcomes. Practically, for encoders that depend only on the input, this is equivalent to penalising input dependence while encouraging task-relevant information, in other words, reducing $I(X;Z)$ while increasing $I(Y;Z)$ (see Methods). The second inequality ensures that predictive power is maintained within a tolerance $\varepsilon$, representing the marginal loss of fidelity inherent to aggressive compression. While this formally recovers the classical Information Bottleneck objective~\cite{tishby2000}, our framework identifies a critical biological constraint. Unlike artificial systems that can optimise these terms simultaneously, survival-critical systems must decouple them. This necessitates a temporal implementation of the bottleneck, optimising the trade-off not through a single loss function but through a sequential wake-sleep cycle, see below.

Defining the predictive target $Y$ requires varying interpretations depending on the ecological context. In supervised tasks, $Y$ corresponds to an explicit label. However, in observational or episodic settings lacking explicit supervision, we define $Y$ as the \textit{predictive target}, typically the successor state $X_{t+1}$. Information in the current episode that does not predict the future is, by definition, noise. This aligns with the successor representation framework where memory is compressed to encode expected future occupancy rather than current sensory details~\cite{stachenfeld2017hippocampus}.

\subsection{Offline Refinement as a Generalisation Filter}
\label{sec:why-offline}

Single-pass encoding faces a fundamental conflict. The rapid acquisition of new information pushes representational complexity $I(X;Z)$ upwards to capture detail necessary for immediate survival. Generalisation, however, demands the opposite. To resolve this tension, we distinguish two distinct phases of learning that separate these objectives temporally:

\begin{enumerate}[leftmargin=1.25em,itemsep=0.5em]
    \item \textbf{Encoding / Perception (Online):} During wakefulness, the system must minimise sensory prediction error on the \textit{current} input. This drives the representation $Z$ to be high-fidelity (high $I(X;Z)$) to support immediate inference. In predictive coding~\cite{rao1999predictive,friston2005theory}, this corresponds to the iterative settling of the state $Z$ to match the sensory clamp $X$.

    \item \textbf{Consolidation (Offline):} During sleep or rest (cf.\ the wake-sleep algorithm~\cite{hinton1995wake}), the system disconnects from sensory input. The objective shifts from matching $X$ to minimising the description length of the stored trace $Z$ conditional on the predictive target $Y$. 
    The consolidation operator $\mathcal{C}$ actively compresses stored traces by minimising their description length conditional on the predictive target $Y$, discarding nuisance detail while preserving outcome-relevant structure (see Methods for objective functions). The Data-Processing Inequality (DPI)~\cite{cover2006elements} guarantees that this refinement, operating without re-accessing sensory input, cannot increase $I(X;Z\mid Y)$, ensuring monotonic convergence toward a more compressed representation. The same principle applies at retrieval: when a stored trace is accessed in a new context, it can undergo further compression informed by the current computational state~\cite{dudai2004, hahamy2023human, bridge2012neural}.
  
\end{enumerate}

This temporal separation explains why biological systems exhibit prolonged, multi-stage consolidation. However, a theoretical danger remains: if the offline process is unconstrained, a high-capacity downstream readout (parameters $\Omega$) could still overfit by memorising residual noise in the consolidated trace. We propose that replay resolves this because the readout depends on the training set $S$ only through the representations $Z$, forming the Markov chain $S \to Z \to \Omega$ (Fig.~\ref{fig:intro}c). By the DPI, the readout cannot acquire more information about $S$ than is present in the representations it accesses ($I(\Omega;S)\le I(Z;S)$). Training on compressed traces $Z_T$ rather than raw encodings $Z_0$ therefore prevents downstream circuits from memorising input-specific noise, regardless of their architectural capacity. In this sense, replay implements a generalisation filter on downstream learning.

\subsection{The Computational Necessity of Temporal Separation}
\label{sec:necessity-separation}

A fundamental question for consolidation theories is whether the same generalisation benefits could be achieved purely online. Why not simply employ stronger regularisation tools (e.g., Variational Information Bottleneck~\cite{alemi2016deep}, high dropout~\cite{srivastava2014dropout,gal2016dropout}) or deeper encoders $X \to Z \to Z' \to Z'' \to Y$ during the initial learning phase? We argue that temporal separation---encoding first, compressing later---is a requirement for resolving the conflict between fidelity and generalisation in high-capacity systems.

\emph{The Fidelity-Generalisation Conflict.} 
The online acquisition phase is governed by the imperative of immediate perception: the system must capture the current input $X$ with sufficient fidelity to minimise prediction error (maximising $I(X;Z)$). Generalisation, conversely, requires discarding all information not predictive of the target (minimising $I(X;Z \mid Y)$). Attempting to achieve optimal compression \emph{during} the single-pass encoding of a novel, high-dimensional experience forces a destructive compromise: strong online regularisation effectively "blinds" the system, preventing the capture of detailed episodic features required for one-shot learning. Temporal separation resolves this by allowing the system to operate in two distinct regimes: a high-fidelity accumulation phase (Wake), in which the hippocampus captures detailed episodic traces, followed by a compression phase (Sleep), in which these traces are distilled into generalisation-ready neocortical representations. (See Methods for formal derivation of this conflict).

\emph{Dynamic State Refinement vs. Static Compression.} 
Standard applications of the Information Bottleneck train a fixed encoder $\Phi$ to compress the entire data distribution. However, learning such a universal compression function online requires massive i.i.d.\ data, and continually updating the encoder risks overwriting the high-fidelity encodings needed for one-shot learning~\cite{van2020brain}. This would also introduce the challenge of maintaining alignment between hippocampal traces and evolving neocortical representations~\cite{kali2004off}. Our framework shifts the optimisation target: instead of compressing the parameters of the perception model ($\Phi$), consolidation compresses the specific latent state of the memory ($z$). Crucially, the consolidation operator is a learned function that generalises across inputs. Once trained, it maps any initial encoding $z_0 = g_\Phi(x)$ to a compressed state $z_T$, so that re-encountering the same input produces the compressed representation through the full pathway $x \to g_\Phi(x) \to \psi_\Psi(z_0) \to z_T$ without modifying the encoder. Thus, the consolidator can be thought of as providing neocortical post-processing for the encoder (or pre-processing for the decoder). It is computationally cheaper and more flexible to refine a single stored memory trace post-hoc (optimising the vector $z$ through iterative updates) than to train a perfect encoder that generalises zero-shot to all future inputs.

\emph{Memory Targeting.}
In systems with explicit memory stores (the human hippocampus, fixed Key-Value caches, episodic buffers), the encoder's job effectively ends once the trace is written. The encoder cannot retroactively modify a stored memory $M$. No amount of encoder depth or online regularisation can compress a memory that has already been consolidated into a buffer; only an offline process that retrieves and refines the stored state can optimise the representation after the fact.

\emph{Parameter Efficiency.}
Increasing encoder depth to achieve compression typically inflates the encoder-complexity term $C$ in Eq.~\ref{eq:ib-bound}. This is counter-productive, as it tightens the bound on the data side but loosens it on the parameter side. Offline refinement reduces $I(X;Z \mid Y)$ without inflating the permanent parameter count of the online perception model, avoiding the complexity penalty inherent to deep online encoders.

\section{Implementations Across Systems}

To validate the principle of predictive forgetting, we translate the abstract operator $\mathcal{C}$ into a concrete neural mechanism. Unlike deep feedforward networks (e.g., stacked autoencoders) that process inputs spatially, we model consolidation as a \emph{temporal iteration on stored states}. This process follows a cyclical dynamic: a memory trace is retrieved, processed via internal refinement loops to isolate its predictive core (analogous to ``dreaming'' in generative world models~\cite{Hafner2020Dream, hinton1995wake}) and updated with stronger internal priors (analogous to synaptic homeostasis~\cite{tononi2014sleep}) to reduce representational complexity. We implement this cycle across three distinct architectures: (i) a minimal cortical model that performs offline refinement of latent codes, (ii) biologically plausible predictive coding networks that utilise explicit generative replay, and (iii) Transformer-based LLMs with key-value cache consolidation. Despite architectural differences, consolidation steps in each model both reduce $I(X;Z)$ and increase $I(Y;Z)$, implementing the optimal generalisation strategy.

\subsection{Consolidation as Offline Refinement of Cortical Latent Codes}
\label{sec:ae-consolidation}

We first model systems consolidation with a minimal abstraction, a cortical circuit that performs \emph{offline refinement} of internal codes without re-accessing raw sensory input. We utilise a deterministic perceptual encoder $g_\Phi: X\to Z$ that is trained unsupervised and then \emph{frozen}. This mirrors the biological observation that early sensory cortices stabilise largely during development to provide a consistent vocabulary for downstream areas~\cite{hensch2005critical}. A downstream readout network $f_\Omega: Z\to Y$ (a shallow multi-layer perceptron parametrised by $\Omega$) is trained to decode these representations into behavioural decisions.

Consolidation is implemented by iteratively updating the latent state using a lightweight refiner $\psi_\Psi$ whose parameters are shared across consolidation steps. Crucially, this refiner operates solely in the latent space ($z \leftarrow z + \psi_\Psi(z)$), refining the initial online encoding $z_0$ without re-accessing the original input $x$ (see Methods for objective functions). The refiner's parameters are trained using task outcomes $Y$ via a frozen readout network, but once trained, each consolidation step requires only the stored latent state.

\begin{figure}[t]
    \centering
    \includegraphics[width=0.9\linewidth]{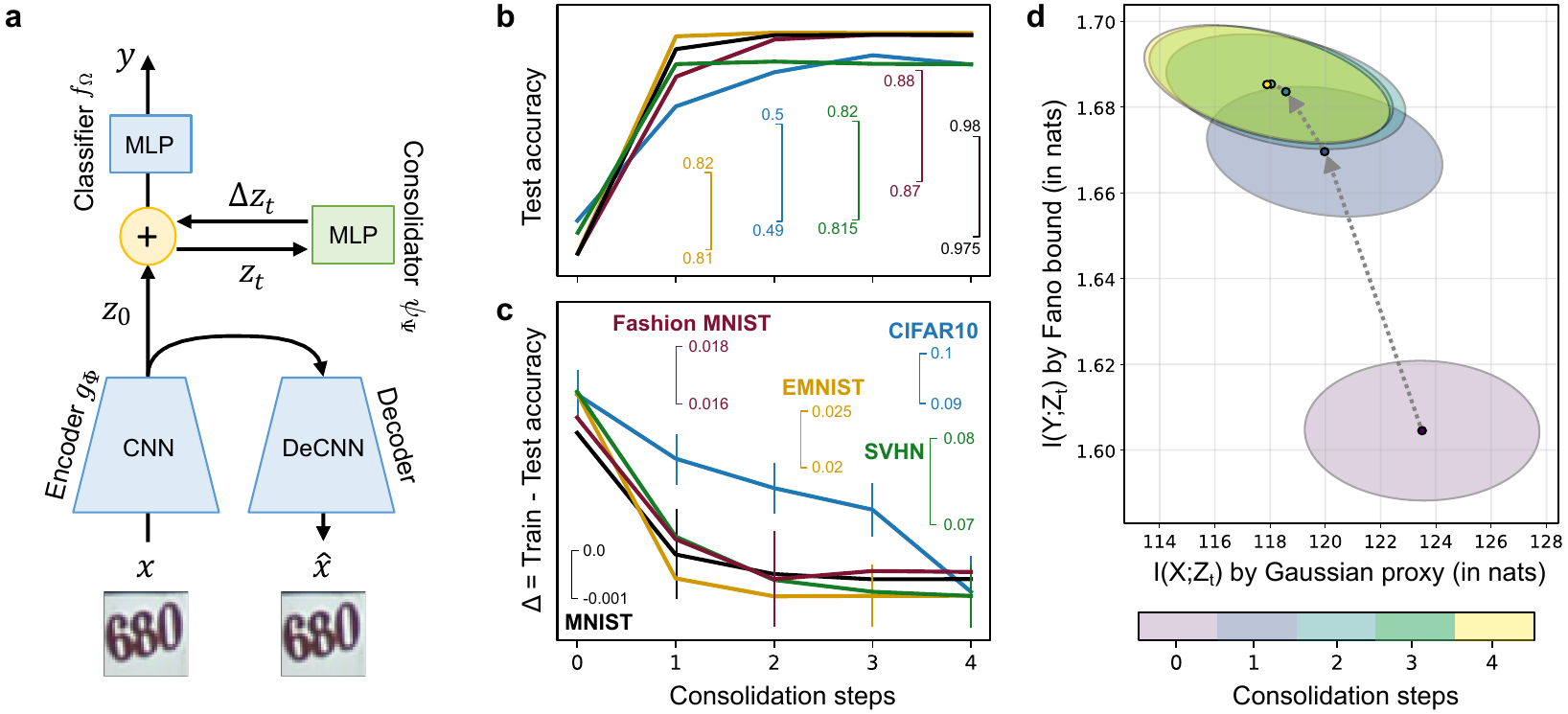}
    \caption{\textbf{Iterative refinement tightens the generalisation bound.} 
    \textbf{a,} Architecture comprising a frozen convolutional encoder, an iterative latent refiner, and a task-dependent readout network. 
    \textbf{b-c,} Effect of consolidation steps on classification accuracy (b) and the generalisation gap $\Delta$ (c) across five image classification benchmarks 
(MNIST, Fashion-MNIST, EMNIST, CIFAR-10, SVHN). Refinement consistently improves performance while shrinking $\Delta$.
    Online-only regularisation baselines defining the fidelity-generalisation frontier for MNIST are shown in Extended Data Fig.~\ref{fig:s1}.
    \textbf{d,} Information-theoretic validation (information measured in nats, i.e.\ natural units). We computed proxies for the mutual information terms in Equation~\ref{eq:ib-bound}. As predicted, consolidation reduces superfluous input dependence $I(X;Z_t)$ (gray) while increasing task-relevant information $I(Y;Z_t)$ (blue), confirming that the system implements predictive forgetting.}
    \label{fig:results:ae}
\end{figure}

To ensure that consolidation improves generalisation through compression rather than memorisation, we employed a \emph{cross-fit} strategy. We partitioned the training data into two disjoint sets: Set A was used exclusively to train the readout network $\Omega$, while Set B was used to train the refiner $\psi_\Psi$. Throughout this process, the underlying perceptual encoder $g_\Phi$ remained frozen. This strict separation prevents the refinement process from exploiting task supervision to memorise the specific examples used by the readout, thereby keeping the encoder-complexity term $C$ fixed (Eq.~\ref{eq:ib-bound}), while using cross-fit to minimise leakage and co-adaptation between refinement and readout. While this strict separation is an experimental control, it mirrors the biological dissociation between rapid hippocampal encoding and slower neocortical consolidation, where distinct neural populations and timescales govern each process. The refiner was trained with a halting objective to minimise the number of computational steps required for the readout to reach a confident prediction of the task outcome $Y$, implementing a ``resource-rational'' consolidation process.

We evaluated the model across five visual classification datasets. To verify the theoretical predictions, we quantified two information-theoretic proxies: (i) a Fano lower bound on task-relevant information $I(Y;Z_T)$, and (ii) a Gaussian proxy for input dependence $I(X;Z_T)$ (see Methods). The results are consistent with the predictive forgetting hypothesis. As shown in Figure~\ref{fig:results:ae}D, iterative offline refinement progressively reduced the superfluous information $I(X;Z_T)$ monotonically with each step $t$, while simultaneously increasing the predictive information $I(Y;Z_T)$. Crucially, this information-theoretic compression translated directly to behavioural outcomes: the consolidated representations exhibited a significantly narrower train--test generalisation gap compared to the single-pass baseline (Figure~\ref{fig:results:ae}C). 

Furthermore, we tested whether these gains could be matched by strengthening single-pass online regularisation. Across a broad sweep of Variational Information Bottleneck and dropout settings, tighter regularisation reduced the generalisation gap but consistently incurred an accuracy cost, tracing a fidelity–generalisation frontier (Extended Data Fig.~\ref{fig:s1}). Offline refinement breaks this trade-off: it achieves a similarly tight gap without sacrificing task performance. Notably, this improvement arises from the replay/refinement mechanism itself rather than delicate hyperparameter tuning, providing a minimal existence proof that iterative operations on stored latent codes—without re-accessing sensory input—can directly tighten the generalisation bound.

\subsection{Biologically Plausible Compression via Bidirectional Predictive Coding}
\label{sec:pc-consolidation}

We next implemented the principle of predictive forgetting using a biologically plausible architecture: hierarchical predictive coding (PC) networks. Unlike autoencoders, PC circuits perform inference through the reciprocal exchange of prediction errors and expectations~\cite{rao1999predictive,friston2005theory}. Recent theoretical work highlights the inherent bidirectionality of these circuits: the same generative weights used for perception can be driven top-down to support imagery and dreaming~\cite{oliviers2025bidirectional}.

\begin{figure}[t]
    \centering
    \includegraphics[width=0.8\linewidth]{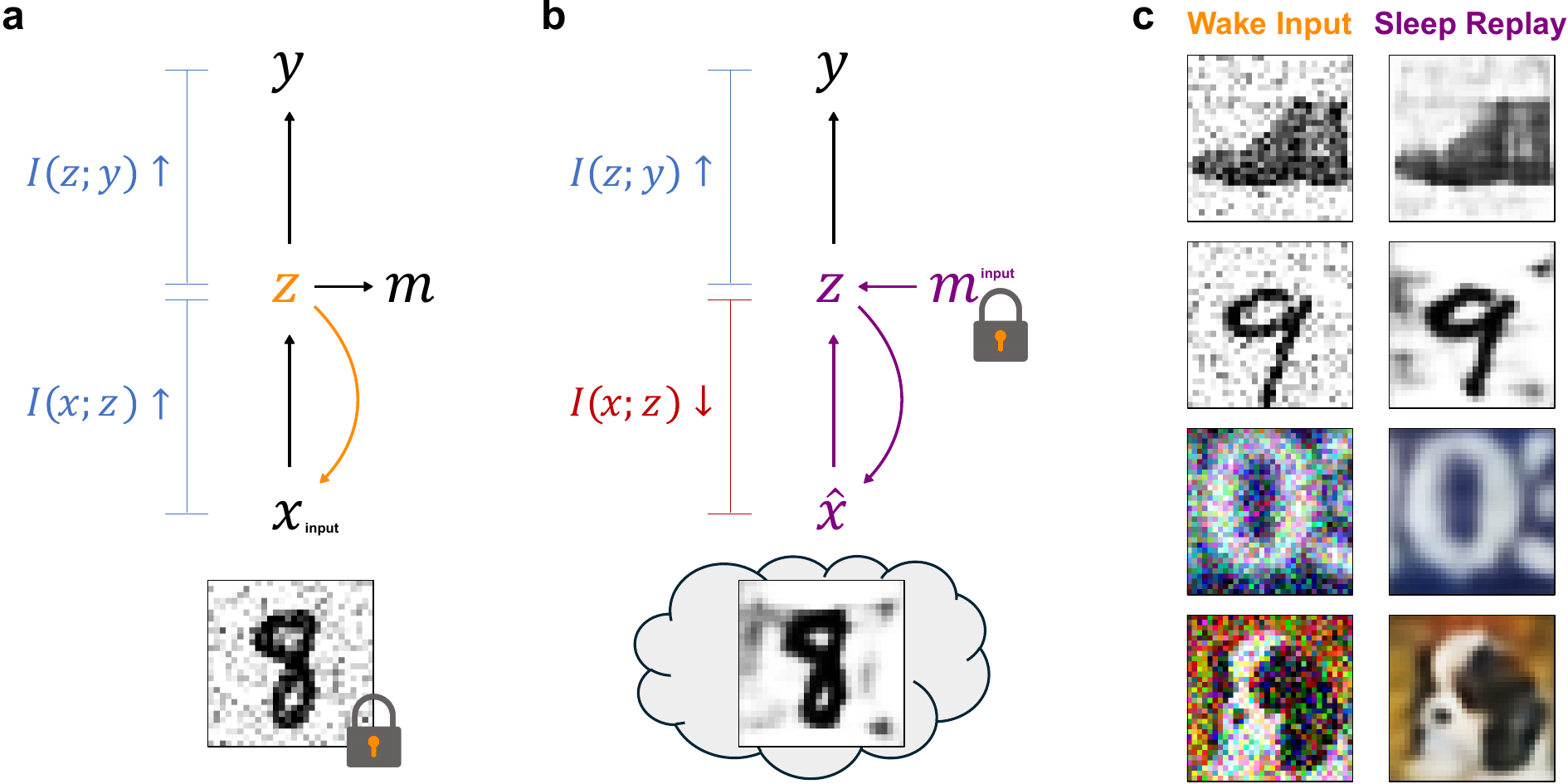}
    \caption{\textbf{The Bidirectional Consolidation Mechanism.}
    \textbf{a, Wake Perception:} Sensory input $x$ is clamped (Lock icon), driving bottom-up inference. The network balances sensory fidelity against internal priors, resulting in a high-fidelity representation where input information $I(x;z)$ is high (Orange).
    \textbf{b, Sleep Consolidation:} Sensory units are unclamped. A stored memory trace $m$ (representing stable synaptic weights or buffer entries) is retrieved to initialise the transient neural activity $z$. The generative model projects this trace top-down to create a "dream" $\hat{x}$ (Cloud icon), which then drives a second pass of precision-weighted inference. The strong homeostatic prior actively minimises representational cost ($I(x;z) \downarrow$, Red), effectively denoising the trace while preserving task-relevant information ($I(z;y)$, Blue).
    \textbf{c, Qualitative Results:} Comparison of noisy wake inputs (Left) vs. consolidated sleep replays (Right) across four datasets. The generative loop filters out high-frequency sensory noise, extracting the semantic gist.}
    \label{fig:dreams}
\end{figure}

We propose that consolidation exploits this bidirectionality to implement a self-correction mechanism. The network consists of a convolutional encoder-decoder (matching the architecture in Section~\ref{sec:ae-consolidation}) with a $d$-dimensional latent layer, where $d$ is varied across experiments to test capacity dependence. To rigorously isolate the effect of consolidation from computational capacity, we matched the inference depth and noise magnitude across both phases (see Methods), varying only the source of information and the strength of the priors. The generative model weights are frozen throughout this experiment to isolate the effect of the inference dynamics on representational compression (downstream readout learning is examined in Section~\ref{sec:replay-necessity}; see Figure~\ref{fig:dreams}a-b for a schematic).

\begin{enumerate}[leftmargin=1.25em,itemsep=0.25em]
    \item \textbf{Wake Perception (Inference).} Sensory input $x$ is clamped. The network performs inference by iteratively updating latent states $z$ via gradient descent on the variational free energy (connection weights are fixed during this phase; see Methods). This balances bottom-up sensory prediction error against top-down prior constraints, driving the network into a high-variance approximate posterior that captures input-specific noise.
    \item \textbf{Sleep Consolidation (Generative Replay).} The sensory units are unclamped. A stored trace $m$ (from the wake phase) is projected top-down to generate a ``dream'' $\hat{x} = \mathbb{E}[p(x \mid z=m)]$. This dream represents the system's \textit{platonic ideal} of the memory, effectively filtering out stochastic noise that the generative model cannot explain. The dream $\hat{x}$ is then clamped as input and the network performs a second pass of inference under stronger homeostatic constraints~\cite{tononi2014sleep}, implemented by increasing the prior weight in the free energy to penalise large latent activations. This forces the representation to settle into a more compressed state $z_{\text{refined}}$ within the same number of inference steps.
\end{enumerate}

\subsection{The Necessity of Replay: Constraining Readout Complexity}
\label{sec:replay-necessity}

\begin{figure}[t]
    \centering
    \includegraphics[width=\linewidth]{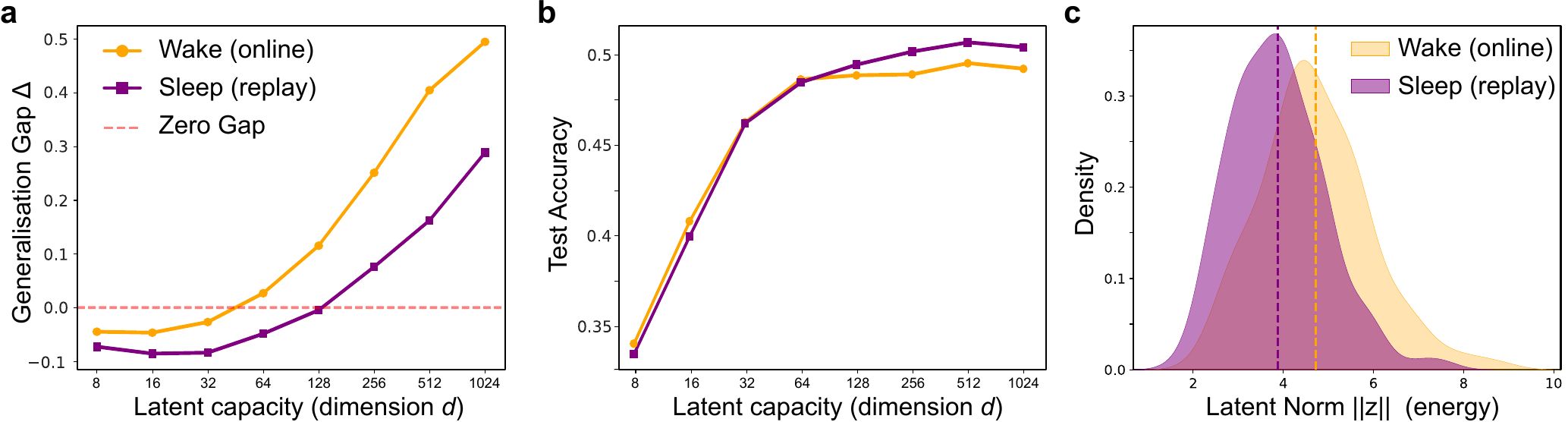}
    \caption{\textbf{Consolidation resolves the capacity-generalisation trade-off.}    
    \textbf{a,} Generalisation Gap ($\Delta$) vs. Representational Capacity ($d$) on CIFAR-10. In low-capacity regimes ($d < 64$), architectural bottlenecks force such strong compression that models underfit training noise, resulting a negligible or negative gap (driven by regularisation). However, as capacity scales ($d \to 1024$), the Online agent (Orange) suffers catastrophic overfitting, memorising sensory noise. Offline Replay (Purple) drastically reduces this gap, allowing the system to scale to high capacities with a significantly minimised generalisation penalty.
    \textbf{b,} Test Accuracy. In low-capacity networks, replay provides minimal benefit as the architecture itself enforces compression. In high-capacity regimes ($d \ge 256$), characteristic of mammalian neocortex, replay translates into a net performance gain. The Replay agent (Purple) outperforms the Online agent because the sleep phase applies stronger internal constraints (priors) that are unavailable during rapid online perception. This active filtering removes the input-specific noise that high-capacity networks otherwise memorise, allowing the system to leverage its full capacity for semantic discrimination. 
    \textbf{c,} Mechanism of Compression ($d=512$). Latent norm distributions reveal the physical basis of this effect: strong homeostatic pressure during sleep contracts memory traces toward the generative manifold (Purple), explicitly reducing representational cost $I(X;Z)$ compared to the high-entropy wake state (Orange).}
    \label{fig:capacity_sweep}
\end{figure}

While the wake-sleep cycle described above compresses individual representations, our theoretical analysis (Section~\ref{sec:why-offline}) predicts that a key benefit of offline consolidation lies in constraining the information available to the downstream readout ($\Omega$), by training on compressed traces rather than raw encodings. We tested this prediction by comparing two agents: an ``Online'' agent trained on raw wake traces, and a ``Replay'' agent trained on consolidated traces, across a logarithmic sweep of representational capacities (Figure~\ref{fig:capacity_sweep}A-B).

We found that the benefits of consolidation scaled with network capacity. In low-capacity regimes ($d < 64$), the architectural bottleneck naturally enforces compression, rendering replay redundant. However, in high-capacity regimes ($d \ge 256$), which more closely mimic the massive synaptic availability of the neocortex, the Online agent (Figure~\ref{fig:capacity_sweep}, Orange) utilises its excess capacity to memorise sensory noise, resulting in a widening generalisation gap.

In contrast, the Replay agent uses training on consolidated traces (during ``Sleep'') to compress internal representations $Z$. As shown in Figure~\ref{fig:capacity_sweep}C, the stronger prior during sleep reduces the magnitude of latent representations, meaning they occupy a smaller region of the latent space and carry less information about the input. This effectively neutralises the overfitting risk of the high-capacity network. Consequently, the Replay agent not only maintains a tight generalisation gap but also achieves superior test accuracy in the high-capacity regime (Figure~\ref{fig:capacity_sweep}A-B). This provides a normative explanation for why the mammalian neocortex, which possesses immense capacity, requires prolonged offline consolidation to generalise well.

\subsection{Hierarchical and Iterative Refinement in High-Capacity Attentional Systems}
\label{sec:llm-consolidation}

To test the universality of predictive forgetting in systems approaching biological scale, we use LLMs as \textit{in-silico} model organisms. The claim is not architectural homology: rather, LLMs provide a fully observable high-capacity system in which consolidation-like interventions can be diagnosed at a resolution that remains out of reach \textit{in vivo}. Modern Transformer architectures rely on a Key--Value (KV) cache to maintain recent context~\cite{vaswani2017}, where ``Keys'' provide a structural addressing system and ``Values'' carry the associated content~\cite{gershman2025keyvalue}. As initially encoded, each entry is instance-specific, functioning as an episodic record of tokens encountered so far~\cite{fountas2025emllm, Dong:2025:TiCS}. As the cache grows, this episodic detail accumulates task-irrelevant information (high $I(X;Z \mid Y)$), degrading attentional retrieval, much as uncompressed episodic traces would overwhelm a biological memory system.

\begin{figure}[t]
    \centering
    \includegraphics[width=\linewidth]{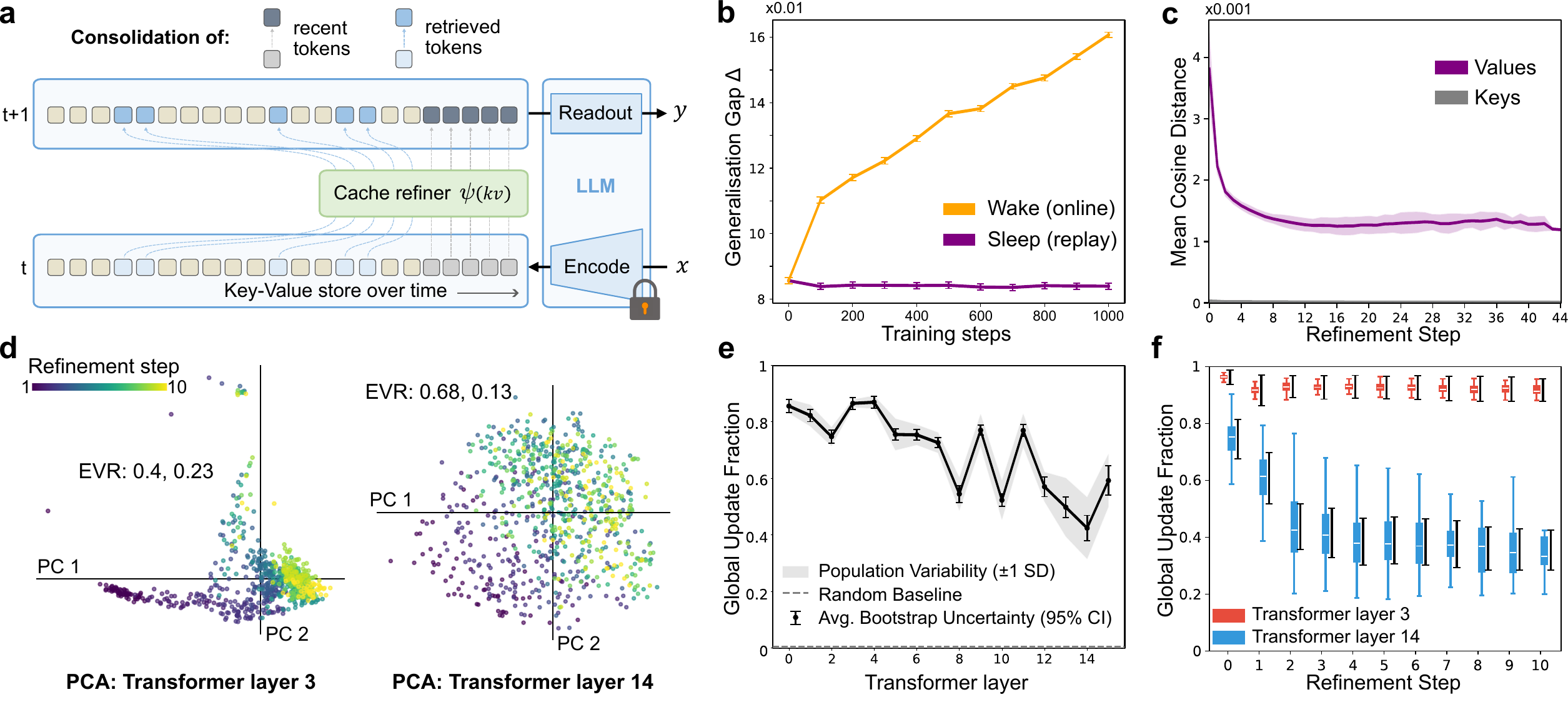} 
    \caption{\textbf{Hierarchical and iterative cache refinement in large language models.}
    \textbf{a,} Schematic of the Bottlenecked Transformer as a model of systems consolidation. A frozen Llama-3-8B backbone (right, lock icon) representing the stable neocortex interacts with a plastic Key-Value store whose entries are iteratively consolidated (left). During offline ``sleep'' phases, a Refiner network ($\psi$) iteratively updates cached representations to minimise conditional mutual information.
    \textbf{b,} Generalisation gap (train vs. test perplexity) on GSM8K reasoning tasks. Offline refinement (\textbf{Purple}) prevents the overfitting observed when fine-tuning the baseline online model (\textbf{Orange}).
    \textbf{c,} Step-wise mean cosine distance, $d(z_t, z_{t+1})$, between cache states before and after each refinement step, for \textbf{Keys (Grey)} and \textbf{Values (Purple)}. Values undergo large initial corrections that diminish but persist across steps, indicating ongoing refinement without convergence to a fixed point. Keys remain nearly unchanged throughout, mirroring the biological distinction between stable structural addressing and malleable content~\cite{gershman2025keyvalue}.
    \textbf{d,} Principal Component Analysis (PCA) of value update trajectories for a single example, coloured by refinement step (Purple=initial $\to$ Yellow=final). Explained Variance Ratios (EVR) for the first two components are shown for each layer. Layer 3 (left) exhibits coherent, parallel flow indicative of Global Renormalisation, whereas Layer 14 (right) shows divergent, complex trajectories indicative of Selective Editing.
    \textbf{e,} Global Update Fraction across layers ($N=1318$). Updates transition from a shared global translation regime ($>0.8$) in early layers to a selective editing regime ($<0.5$) in deep layers. Grey shading indicates population variability ($\pm 1$ s.d.); blue error bars indicate bootstrap uncertainty ($95\% CI$).
    \textbf{f,} Evolution of update coherence over refinement steps. Early layers (\textbf{Red}, Layer 3) maintain global coherence, while deep layers (\textbf{Blue}, Layer 14) progressively become more selective, indicating a coarse-to-fine refinement strategy.}
    \label{fig:llm_results}
\end{figure}

We therefore tested whether offline consolidation could transform these episodic cache entries into compressed representations optimised for prediction, using the Bottlenecked Transformer architecture~\cite{oomerjee:2025:bt} on complex mathematical reasoning tasks (GSM8K). This system augments a pre-trained Transformer model (Llama-3-8B) acting as a frozen backbone with a lightweight ``Cache Refiner'' that iteratively rewrites stored traces during offline intervals and retrieval events (Figure~\ref{fig:llm_results}a). The consolidated cache thus functions analogously to neocortical semantic memory (with veridical retrieval mechanisms such as retrieval-augmented generation~\cite{lewis2020rag} serving a complementary, hippocampal-like role~\cite{Spens2025HippocamponeocorticalIA}). Improving the efficiency of KV cache storage is also a pressing engineering challenge as context windows scale~\cite{oomerjee:2025:bt}.

Consistent with our theoretical predictions, offline refinement fundamentally altered the system's learning dynamics. While the standard online model exhibited progressive overfitting (widening generalisation gap), the consolidated system maintained a tight generalisation bound (Figure~\ref{fig:llm_results}b), effectively closing the gap between training and validation perplexity. Because we have full observability of the cache updates in this system, we were able to dissect the mechanism of consolidation across network layers.

This analysis revealed two key findings. First, consolidation operates differently at different levels of the processing hierarchy. Early layers (0--4) apply a shared, uniform correction across all memory tokens (``Global Renormalisation''), consistent with broad denoising. Deep layers (8--15) instead apply token-specific updates (``Selective Editing''), consistent with fine-grained semantic adjustment (Figure~\ref{fig:llm_results}d-f; see Methods for formal quantification). This coarse-to-fine strategy also unfolds temporally: initial refinement steps are dominated by global corrections, with selective editing emerging progressively (Figure~\ref{fig:llm_results}f).

Second, we observed a functional dissociation between the structural scaffold of memory and its content. Throughout refinement, the geometry of Keys remained stable between consecutive steps, whereas Values exhibited sustained representational change that diminished but never ceased (Figure~\ref{fig:llm_results}c). This mirrors the distinction proposed by Gershman et al.~\cite{gershman2025keyvalue}, where hippocampal keys provide a stable addressing system optimised for discriminability whilst neocortical values store content subject to consolidation. Our results provide the first direct observation of this dissociation during an active compression process.

\section{Testable Predictions and Empirical Validation}

Our framework generates quantitative signatures of how neural representations must change during consolidation. Below, we highlight empirical findings that are consistent with the framework (grounding/unification) and outline distinct quantitative predictions that remain to be tested. Table~\ref{tab:predictions} summarises these signatures.

\begin{table}[h]
\centering
\caption{Predictions and Empirical Support for Predictive Forgetting}
\label{tab:predictions}
\scriptsize
\begin{tabularx}{\textwidth}{p{1.6cm} X p{2.8cm} p{1.3cm} p{2.2cm}}
\toprule
\textbf{Mechanism} & \textbf{Key Neural Signature} & \textbf{Methods to Test} & \textbf{Status} & \textbf{Evidence} \\
\midrule
\textbf{Temporal}\newline \textbf{Compression} & 
- Manifold radius/dimension reduction\newline
- Increased within-category similarity & 
fMRI RSA, Manifold Capacity Analysis & 
\textbf{Grounding} & 
\cite{chung2018classification, schapiro2017scirep} \\
\midrule
\textbf{Sleep}\newline \textbf{Acceleration} & 
- Compression ($I(X;Z) \downarrow$) accelerates during sleep vs. wake & 
Sleep manipulation, fMRI reactivation & 
\textbf{Grounding} & 
\cite{igloi2015, ellenbogen2007human} \\
\midrule
\textbf{Generalisation}\newline \textbf{Link} & 
- Neural compression predicts OOD performance\newline
- Selective for task-irrelevant features & 
Behavioural transfer tests, individual differences & 
\textbf{Grounding} & 
\cite{wimmer2020episodic, tompary2017} \\
\midrule
\textbf{Readout}\newline \textbf{Sharpening} & 
- Increased SNR in memory traces\newline
- Synaptic down-selection of noise & 
Longitudinal 7T fMRI, E-phys & 
\textbf{Grounding} & 
\cite{vanasse2022multiple, devivo2017ultrastructural} \\
\midrule
\textbf{Predictive}\newline \textbf{Codes} & 
- Shift from retrospective to prospective Successor Representations & 
Sequential learning tasks & 
\textbf{Unification} & 
\cite{stachenfeld2017hippocampus, momennejad2017successor, schapiro2017scirep, bennett2025subicular} \\
\midrule
\textbf{Hierarchical}\newline \textbf{Gradients} & 
- Compression magnitude scales linearly with cortical hierarchy & 
Multi-region fMRI & 
\textbf{Prediction} & 
\textit{Partial:} \cite{bhandari2025task, krenz2023} \\
\midrule
\textbf{Reconsolidation} & 
- Retrieval induces plasticity that further compresses trace & 
Retrieval-practice paradigms & 
\textbf{Prediction} & 
\textit{Partial:} \cite{hahamy2023human, bridge2012neural} \\
\midrule
\textbf{Structure/}\newline \textbf{Content Split} & 
- Temporal keys ($K$) stable, sensory values ($V$) compressed & 
LLM analysis / HC subfield recording & 
\textbf{Prediction} & 
\textit{Theory:} \cite{gershman2025keyvalue}; Fig.~\ref{fig:llm_results}\\
\bottomrule
\end{tabularx}
\end{table}

\subsection{Temporal Compression of Neural Manifolds}
\label{sec:p1}
If consolidation implements predictive forgetting, neural representations must become progressively more compressed over time. Geometrically, this corresponds to a reduction in the \emph{radius} and \emph{intrinsic dimension} of the neural manifold~\cite{chung2018classification}. Our theory predicts that pattern similarity within task-relevant categories increases, whilst between-category distinctions are sharpened. This explains the "semanticisation" observed in longitudinal fMRI~\cite{schapiro2017scirep}, reinterpreting it not as abstract knowledge accumulation, but as the active pruning of manifold variance orthogonal to the task.

\subsection{Sleep-Dependent Acceleration}
\label{sec:p2}
If offline consolidation is computationally necessary, sleep should accelerate representational change relative to equivalent wake intervals. Measuring representational geometry immediately post-encoding vs. post-sleep should reveal greater compression (manifold contraction) following sleep. This effect has been behaviourally indexed as "gist extraction"~\cite{igloi2015} and "relational integration"~\cite{ellenbogen2007human}, but our framework offers a precise neural definition: sleep replay selectively samples traces to minimise description length. If neural replay is disrupted, we predict a specific failure in compression (retention of noise), leading to overfitting.

\subsection{Generalisation-Compression Correlation}
\label{sec:p3}
Across individuals and time points, greater neural compression should predict better generalisation to novel exemplars. This provides the critical link between our information-theoretic framework and behavioural outcomes: if reducing $I(X;Z \mid Y)$ tightens generalisation bounds, then participants exhibiting stronger compression should show superior out-of-distribution performance. Evidence regarding the integration of overlapping memories supports this view, showing that neural overlap (compression) predicts inductive inference performance~\cite{tompary2017,wimmer2020episodic}. Importantly, we predict this relationship is selective: compression of information orthogonal to task demands improves generalisation, whereas compression of task-relevant structure impairs it.

\subsection{Readout Sharpening via Competitive Forgetting}
\label{sec:p4}
A critical implication of our framework is that consolidation optimises the decision boundaries of downstream readouts. We posit that this mechanism explains the phenomenon of "trace sharpening" recently observed in high-field fMRI~\cite{vanasse2022multiple}, where the signal-to-noise ratio (SNR) of memory representations increases over weeks. Our model interprets this not as passive decay, but as an active process driven by \emph{synaptic down-selection}: synapses encoding non-predictive nuisance variables are pruned during sleep-dependent renormalisation~\cite{devivo2017ultrastructural}, effectively implementing the information bottleneck.

\subsection{Emergence of Predictive Codes}
\label{sec:p5}
In sequential domains, minimising conditional information ($I(X;Z \mid Y)$) naturally leads to representations that encode expected future state occupancy rather than immediate sensory features. The Successor Representation (SR)~\cite{stachenfeld2017hippocampus} provides a concrete example, where SR-like codes in hippocampal CA1 may reflect the earliest stages of this compression, while the eigenvectors of the SR correspond to grid-like structural codes in entorhinal and prefrontal cortices~\cite{stachenfeld2017hippocampus, whittington2020tolman, behrens2018cognitive}. We predict that consolidation drives a qualitative shift from environment-specific, place-like representations toward these structural, environment-general codes over days to weeks. Notably, grid cells reside at the hippocampal-neocortical interface, consistent with the view that predictive compression propagates from hippocampal episodic traces toward neocortical structural representations~\cite{spens2024generative, bennett2025subicular}.

\subsection{Hierarchical Gradients of Compression}
\label{sec:p6}
We further predict that consolidation-related compression varies systematically across the cortical hierarchy. Early sensory regions, which must preserve input fidelity for immediate perceptual demands, should exhibit minimal compression. Higher associative and parahippocampal cortices (e.g., prefrontal, parietal, and entorhinal regions) that extract abstract task structure should show maximal compression. This predicts a spatial gradient in the magnitude of $I(X;Z \mid Y)$ reduction. Recent evidence of "task-tailored" geometry in prefrontal cortex~\cite{bhandari2025task} and semantic transformation in the hippocampal-neocortical loop~\cite{krenz2023} provides initial support for this view, but the longitudinal evolution of this gradient remains a distinct testable signature of our framework.

\subsection{Reconsolidation and Retrieval-Induced Compression}
\label{sec:p7}
Each memory retrieval event could trigger further compression. Our framework predicts that the inverse loop (recall followed by re-encoding) implements an additional consolidation step. Thus, retrieved memories should show greater compression than non-retrieved memories matched for age. Empirically, this aligns with evidence that retrieval induces plasticity that "updates" the memory trace towards the current context~\cite{hahamy2023human,bridge2012neural}. We further predict this is context-dependent: when task demands require preservation of episodic detail, retrieval-induced compression is suppressed; when abstraction is required, retrieval accelerates the collapse of nuisance dimensions.

\subsection{Structure-Content Dissociation}
\label{sec:p8}
Finally, our model makes a novel mechanistic prediction regarding the internal structure of the memory trace. Based on our LLM results (Fig.~\ref{fig:llm_results}c-d), we predict a physiological dissociation between addressing and content. 
In predictive coding accounts of cortical processing, superficial layers preferentially carry bottom-up prediction errors whilst deep layers convey top-down predictions~\cite{bastos2012canonical, shipp2016neural}. By analogy, neural populations encoding temporal or contextual indices (Keys), potentially in CA3 or superficial cortical layers, should exhibit high stability during consolidation to preserve structural scaffolding. In contrast, populations encoding sensory content (Values), such as those potentially in CA1 or deep cortical outputs, should undergo more aggressive compression. Testing this prediction requires dissecting hippocampal-neocortical codes into address-like and content-like components, offering a roadmap for high-density electrophysiology to distinguish ``indexing'' stability from ``content'' plasticity.

\section{Discussion}

Our information-theoretic framework provides a computational answer to the purpose of memory consolidation: it implements the optimal solution to the generalisation problem. We propose that the brain implements predictive forgetting, the iterative minimisation of $I(X;Z \mid Y)$ while preserving $I(Y;Z)$, which provably tightens generalisation bounds in encoder-decoder architectures. The \emph{fidelity-generalisation conflict} prevents single-pass encoders from achieving this optimum. Fast acquisition necessarily prioritises information capture, whereas optimal generalisation demands compression. Temporal separation resolves this tension, with offline consolidation iteratively discarding superfluous input dependence without sacrificing predictive power. This explains why biological systems exhibit prolonged, multi-stage consolidation during sleep and why replay is computationally essential rather than epiphenomenal.

This account operates at Marr's computational level~\cite{marr2010vision}, specifying what consolidation achieves (optimal generalisation via conditional information compression) rather than merely how it operates. The algorithmic level follows naturally, with iterative refinement via generative replay, hippocampal--neocortical dialogue, or cache reprocessing emerging as implementation strategies for approaching the Information Bottleneck optimum. Finally, the implementation level concerns the specific neural substrates of replay, synaptic plasticity, and systems consolidation.

\subsection{A Unifying Framework for Consolidation}

Normative work in control theory has shown that complexity reduction is essential for generalisation~\cite{moskovitz2024understanding}. Our information-theoretic characterisation establishes an analogous principle for memory, unifying disparate phenomena under a common optimisation objective. Systems consolidation, understood as the gradual memory shift from hippocampal to neocortical dependence~\cite{marr1971simple, mcclelland1995, squire1995}, emerges naturally: as the neocortical generative model accumulates predictive structure, it becomes increasingly self-sufficient for task-relevant inference, while the hippocampus retains veridical episodic details not captured by this compression~\cite{spens2024generative, nicholas2026episodic}. 
This implies that the hippocampus stores bound representations linking sensory inputs, latent states, and behavioural outcomes~\cite{spens2024generative}, providing the associations necessary for offline training of the consolidation operator. Consistent with this, posterior hippocampal representations have been linked to perceptual detail while anterior representations encode more schematic information~\cite{poppenk2013long, sekeres2024}, suggesting a compression gradient within the hippocampal formation itself.
The hippocampus initially encodes detailed, high-$I(X;Z)$ representations necessary for rapid learning. Offline replay progressively refines these traces, reducing $I(X;Z \mid Y)$ through iterative application of the predictive coding update rule. As the neocortical model becomes sufficiently complete to support independent pattern completion, the system relies less on hippocampal retrieval, though veridical details may persist hippocampally for episodic recall and future re-encoding. This is not ``transfer'' in the sense of copying information between regions, but rather the natural consequence of neocortical compression rendering simpler retrieval mechanisms viable.

Representational drift, the observation that neural codes change over time even for stable behaviours~\cite{driscoll2017dynamic}, has often been viewed as noise or network instability. Our framework reinterprets drift as ongoing compression. Representations reorganise to reduce $I(X;Z \mid Y)$, explaining why drift can be substantial yet behaviourally silent. Hence, the seeming paradox dissolves: codes change because they are converging toward minimal sufficient statistics, grouping together inputs with identical posteriors $p(y \mid x)$ regardless of their original sensory similarity. Our LLM experiments provide a concrete illustration: Value representations exhibit sustained step-wise change that diminishes but never ceases (Fig.~\ref{fig:llm_results}c), while Keys remain stable, mirroring the biological observation that content representations drift while structural scaffolding is preserved. This suggests that drift is not a failure of memory maintenance but a signature of ongoing optimisation. Notably, drift is also observed within the hippocampus itself~\cite{ziv2013long}, which may reflect both local compression within the hippocampal formation~\cite{schapiro2017scirep} and the need to maintain alignment with evolving neocortical representations~\cite{kali2004off, van2020brain}.

Memory semanticisation, the gradual loss of episodic detail with retention of conceptual gist~\cite{winocur2010}, corresponds precisely to our definition of predictive forgetting. Contextual details orthogonal to task demands, such as when, where, or incidental features, constitute high $I(X;Z \mid Y)$. These dimensions are input-specific but do not improve predictions about behavioural outcomes. 
Consolidation selectively distils the predictive core from these traces into a neocortical representation, discarding dimensions that do not improve generalisation. The result is a division of labour: the neocortex carries predictive structure while the hippocampus retains the episodic details needed for veridical recall~\cite{spens2024generative,nicholas2026episodic}. This process reflects optimisation rather than degradation.

Furthermore, our framework provides insight into what maximally compressed codes should look like. Representations that minimise $I(X;Z \mid Y)$ in sequential domains naturally abstract from environment-specific sensory details to encode transition structure. The Successor Representation (SR)~\cite{stachenfeld2017hippocampus} exemplifies this: SR-like codes in CA1 may represent an early consolidation step from more episodic CA3 traces~\cite{schapiro2017scirep}, while the eigenvectors of the SR correspond to grid-like structural codes in entorhinal and prefrontal cortices~\cite{stachenfeld2017hippocampus, whittington2020tolman}. These neocortical grid codes generalise across environments, consistent with the endpoint of predictive compression: structural representations that capture task-relevant transitions independently of specific sensory contexts.

In artificial systems, continual learning provides a compelling test case. Catastrophic forgetting results from interference between incompressible, task-specific representations. Our framework provides a normative justification for consolidating old representations before learning new ones~\cite{kirkpatrick2017overcoming, van2020brain}: compressed representations occupy less capacity, exhibit greater stability under subsequent learning, and transfer more readily to related tasks. Recent work on surprise-driven replay with dual-rate consolidation demonstrates the benefits of biologically inspired memory management in language models~\cite{Hazard:2025:sure}, though explicit offline compression of stored representations remains to be integrated. More broadly, as LLMs and robotic systems face increasingly open-ended, lifelong learning demands, explicit consolidation phases may prove essential, offering a principled alternative to approaches that rely on regularisation or replay of raw data.

Knowledge distillation in machine learning, which involves training smaller student networks to match the outputs of larger teacher models~\cite{hinton2015distilling}, implements predictive forgetting 
through architectural compression. The student network necessarily has a lower $I(X;Z)$ due to fewer parameters, yet is trained to preserve $I(Y;Z)$ by matching teacher predictions. This forced compression often improves generalisation beyond the teacher, exactly as our framework predicts. Distillation can therefore be viewed as a means to implement consolidation.

Finally, episodes encoded with high fidelity due to surprise undergo more extensive consolidation-related change, explaining why novel memories show greater representational drift during offline periods~\cite{driscoll2017dynamic}. Our previous work on surprise-driven encoding, event segmentation, and memory replay~\cite{fountas2022predictive,fountas2025emllm,mariola2022event,Hazard:2025:sure} addressed the initial encoding phase computationally, whereas the present work focuses on subsequent consolidation. Together, these frameworks suggest that memory compression operates as a two-stage process, consisting of adaptive encoding based on predictability, followed by progressive offline refinement.

\subsection{Limitations and Boundary Conditions}

Our framework rests on several assumptions that warrant examination. The generalisation bound in Equation~\ref{eq:ib-bound} assumes bounded loss functions (i.e., prediction errors that cannot grow arbitrarily large) and deterministic or low-noise encoders to enable clean application of the Data Processing Inequality; extension to unbounded losses or highly stochastic encoders requires additional treatment. While we have partially addressed unsupervised settings by defining $Y$ as the successor state (Section~\ref{sec:consolidation-def}), a comprehensive treatment of purely self-supervised objectives remains an important direction. The bound is asymptotic in sample size, and finite-sample behaviour may differ quantitatively in low-data regimes common in biological learning.

Several phenomena fall outside our current account. Emotionally salient memories may be stored as strong sensory details via amygdala-dependent mechanisms that partially bypass the consolidation process described here~\cite{mcgaugh2000}, and our model does not currently make contact with individual differences in consolidation efficiency or the details of sleep physiology. Paradoxically, consolidation sometimes strengthens false memories or distorts recall toward schematic expectations~\cite{loftus1975}. These are cases in which compression may overshoot optimality, sacrificing accuracy for consistency. Such boundary conditions suggest that consolidation is not driven purely by generalisation but is also shaped by metabolic constraints, social demands, and evolutionary pressures that we have not modelled.

The hierarchical transition from global renormalisation to selective editing observed in our \textit{in-silico} model organism (Section~\ref{sec:llm-consolidation}) would likely remain invisible to current biological assays. Our simulations thus provide a window into representational dynamics that our framework predicts, but are currently measurable only in fully observable artificial systems operating in a high-capacity regime. Our approach is therefore to identify mechanistic signatures in biological data using measurements that are currently feasible (Table~\ref{tab:predictions}).

\subsection{Future Directions}

Longitudinal studies tracking representational changes across encoding, wake retention, and post-sleep intervals can directly test predictions in Sections~\ref{sec:p1}--\ref{sec:p8}. If representational geometry changes as predicted, including decreased dimensionality, increased within-category similarity, and reduced decodability of task-irrelevant features, this will provide direct neural evidence that consolidation implements conditional information compression. Sleep manipulation studies and targeted disruption of replay could further isolate the mechanisms driving compression.

Theoretical extensions include formalising multi-task consolidation, treating stochastic encoders, deriving finite-sample bounds, and incorporating biological constraints such as metabolic costs and temporal discounting. Connecting predictive forgetting to rate-distortion theory~\cite{nagy2025adaptive, shannon1959coding} and PAC-Bayes bounds~\cite{mcallester1999pac, catoni2007pac} would further situate it within the broader landscape of learning theory.

Applications to artificial intelligence may prove the framework's most immediate impact. LLMs facing constraints from growing context windows could implement periodic cache consolidation to maintain performance while reducing memory footprint, and complementary approaches that organise sequences into episodic events~\cite{fountas2025emllm} could benefit from compression mechanisms that preserve relational structure. More broadly, model compression techniques such as pruning, quantisation, and distillation could be unified under the predictive forgetting objective, optimising the compression-relevance trade-off rather than minimising parameters heuristically. If consolidation is indeed the optimal solution to a fundamental computational problem, then artificial systems facing that same problem should benefit from implementing the same principle, regardless of their substrate.

\section{Conclusion}

We have established that memory consolidation can solve a fundamental computational problem: learning representations that generalise beyond training experience. The progressive elimination of statistical dependence between memories and their original inputs, conditioned on behavioural outcomes, which we term predictive forgetting, provably tightens generalisation bounds in encoder-decoder architectures. Critically, this optimisation cannot be achieved through initial encoding alone. The tension between rapid acquisition (which prioritises information capture) and optimal generalisation (which demands compression) necessitates temporal separation. Iterative offline refinement is thus a computational requirement for approaching the information bottleneck optimum.

Our framework unifies many phenomena. Systems consolidation, representational drift, memory semanticisation, successor representations, continual learning stability, and knowledge distillation emerge as manifestations of the same optimisation objective operating in different architectural substrates. Whether implemented through hippocampal--neocortical dialogue during sleep, predictive coding loops in sensory hierarchies, or cache refinement in LLMs, the principle remains constant: reduce $I(X;Z \mid Y)$ whilst preserving $I(Y;Z)$ through iterative processing. This provides the first computational account that explains not merely how consolidation operates, but why it must exist in intelligent systems facing the generalisation problem.

The theory generates quantitative predictions, offering a roadmap for empirical validation. It suggests principled solutions for artificial systems struggling with catastrophic forgetting, context-window limitations, and sample-inefficient learning. Most fundamentally, it answers a longstanding question in neuroscience and artificial intelligence regarding the reason for offline consolidation: in high-capacity systems operating under high-fidelity online demands, consolidation is computationally necessary to approach optimal generalisation, because single-pass encoding retains outcome-irrelevant detail. Predictive forgetting, therefore, identifies a normative target for consolidation and explains why iterative offline refinement is ubiquitous across biological and artificial learners. Consolidation is intelligence’s solution to the problem of generalisable learning from experience.

\section*{Methods}

\subsection*{Theoretical Derivations}

\paragraph{Information-Theoretic Generalisation Bound.}
The bound in Equation~\ref{eq:ib-bound} relies on recent advances in learning theory~\cite{kawaguchi2023}. Let $(X,Y)\sim\mathcal{D}$ be input--target random variables and $S=\{(x_i,y_i)\}_{i=1}^n$ be a training set drawn i.i.d.\ from $\mathcal{D}$. We consider a possibly stochastic encoder mapping $\phi_S: \mathcal{X} \to \mathcal{Z}$, and a decoder $g_S(Z)$. Let $\Phi_S$ denote the random variable representing the parameters of this encoder (dependent on the training set $S$). The generalisation gap $\Delta(S)$ is bounded with high probability by:
\begin{equation}
    \Delta(S) \le \sqrt{\frac{2\sigma^2 \big(I(X;Z \mid Y) + I(\Phi_S;S) + 1\big)}{n}}
\end{equation}
where $I(\Phi_S;S)$ is the mutual information between the encoder parameters and the dataset. In our analysis of consolidation, we assume the encoder is frozen (or slowly changing), making $I(\Phi_S;S)$ constant. Thus, the bound is dominated by the conditional mutual information term $I(X;Z \mid Y)$ and in Eq.~\ref{eq:ib-bound}, we abbreviate $C := I(\Phi_S;S) + 1$.

\paragraph{Decomposition via the Chain Rule.}
We define predictive forgetting as the minimisation of $I(X;Z \mid Y)$. By the chain rule of mutual information:
\begin{equation}
    I(X;Z\mid Y) = I(X;Z) - I(Y;Z) + I(Y;Z\mid X)
\end{equation}
For deterministic or low-noise encoders, $Z$ is a function of $X$, implying $I(Y;Z \mid X) \approx 0$. Therefore, minimising $I(X;Z \mid Y)$ is equivalent to minimising input complexity $I(X;Z)$ while maximising task relevance $I(Y;Z)$, recovering the classical Information Bottleneck objective~\cite{tishby2000}: $\min_{Z} I(X;Z) - \beta\, I(Y;Z)$, which seeks the maximally compressed representation that retains task-relevant information.

\paragraph{Monotonicity via Data Processing Inequality.}
Offline consolidation applies a refinement operator to stored states without re-accessing sensory input. In our implementations, the refinement step acts solely on the current latent state ($Z'=\mathcal{C}(Z)$); the outcome $Y$ is used to train the operator parameters (on the cross-fit split $\mathcal{B}$) but is not provided as an input during refinement. Thus the refinement dynamics form a Markov chain $X \to Z \to Z'$. By the Data Processing Inequality~\cite{cover2006elements}, post-processing cannot add new information about the input, and therefore
\begin{equation}
    I(X;Z' \mid Y) \le I(X;Z \mid Y).
\end{equation}
Iterating $\mathcal{C}$ makes $I(X;Z^{(t)}\mid Y)$ non-increasing. Since Eq.~\ref{eq:ib-bound} is monotone in $I(X;Z\mid Y)$, decreasing this term tightens the corresponding generalisation guarantee; in our implementations this is enforced while preserving task information (e.g., maintaining $I(Y;Z)$ within tolerance as in Eq.~\ref{eq:consolidation}). In high-capacity regimes where the initial encoding $Z$ retains nuisance detail, this typically yields a progressive reduction of input dependence toward a predictive sufficient statistic.

\paragraph{Readout Regularisation via Replay.}
While the encoder complexity $I(\Phi;S)$ is assumed constant, the generalisation capacity of the system is also constrained by the downstream readout mechanism, parameterised by $\Omega$. We model the flow of information to the readout as a Markov chain. During online learning, the readout accesses raw encodings: $S \to Z_0 \to \Omega_{\text{online}}$. During offline replay, it accesses consolidated traces: $S \to Z_0 \to Z_T \to \Omega_{\text{replay}}$.

By the DPI, the information the readout can acquire about the training set $S$ is upper-bounded by the information in the latent representation it accesses. Since consolidation is defined as a lossy compression operation $T$ such that $I(Z_T; S) < I(Z_0; S)$, it follows that:
\begin{equation}
    I(\Omega_{\text{replay}}; S) \le I(Z_T; S) < I(Z_0; S) \approx I(\Omega_{\text{online}}; S)
\end{equation}
This inequality demonstrates that training on consolidated traces physically restricts the search space of the readout parameters $\Omega$, preventing the memorisation of nuisance variables present in $Z_0$ but absent in $Z_T$, regardless of the readout's architectural capacity.

\subsection*{Formal Derivation of the Fidelity-Generalisation Conflict}
\label{sec:proof-conflict}

Here we provide the formal justification for the claim that single-pass encoding cannot simultaneously satisfy the requirements of immediate perception and optimal generalisation in noisy environments.

Let $X$ be the input, $Z$ the representation, and $Y$ the task variable.
\begin{enumerate}
    \item \textbf{The Online Fidelity Constraint.} During initial encoding (Wake), the system must be able to distinguish the current state $X$ from other potential states to enable one-shot learning or immediate physical interaction. Formally, this requires minimising the conditional entropy $H(X \mid Z)$ below some threshold $\epsilon \to 0$, since the system cannot yet distinguish signal from noise. By definition of mutual information $I(X;Z) = H(X) - H(X \mid Z)$, this implies:
    \begin{equation}
        I(X;Z) \approx H(X)
    \end{equation}
    The representation must effectively carry the full entropy of the input.

    \item \textbf{The Generalisation Objective.} As established in Equation~\ref{eq:ib-bound}, optimising the generalisation bound requires minimising the conditional mutual information $I(X;Z \mid Y)$. By the chain rule of mutual information,
    \begin{equation}
        I(X;Z \mid Y) = I(X;Z) - I(Y;Z) + I(Y;Z \mid X).
    \end{equation}
    Under our assumption of deterministic encoders, where $Z=g(X)$ and thus $I(Y;Z \mid X)=0$, this reduces to:
    \begin{equation}
        I(X;Z) = I(X;Z \mid Y) + I(Y;Z).
    \end{equation}
    
    \item \textbf{The Conflict.} Substituting the Fidelity constraint ($I(X;Z) \approx H(X)$) into the decomposition yields:
    \begin{equation}
        H(X) \approx I(X;Z \mid Y) + I(Y;Z)
    \end{equation}
    Rearranging for the generalisation cost $I(X;Z \mid Y)$:
    \begin{equation}
        I(X;Z \mid Y) \approx H(X) - I(Y;Z)
    \end{equation}
    In any realistic high-capacity environment, the input contains nuisance variables (noise) such that the input entropy $H(X)$ is strictly greater than the task-relevant information $I(Y;X)$. Since $I(Y;Z)$ cannot exceed $I(Y;X)$ (DPI), it follows that:
    \begin{equation}
        I(X;Z \mid Y) \ge H(X) - I(Y;X) > 0
    \end{equation}
\end{enumerate}

So long as the system is required to maintain high fidelity ($I(X;Z) \approx H(X)$), the superfluous information $I(X;Z \mid Y)$ is lower-bounded by the noise level of the environment ($H(X \mid Y)$). It is mathematically impossible to drive $I(X;Z \mid Y) \to 0$ (optimal generalisation) while maintaining $I(X;Z) \to H(X)$ (optimal perception).
Temporal separation resolves this by toggling the constraint: at $t=0$ (Wake), we satisfy Fidelity; at $t>0$ (Sleep), we relax Fidelity to minimise $I(X;Z \mid Y)$.

\subsection*{Information-Theoretic Estimators}
To quantify the reduction in conditional mutual information $I(X;Z \mid Y)$ and the preservation of task information $I(Y;Z)$, we employed two proxy measures suitable for high-dimensional latent spaces.

\textbf{Task-Relevant Information ($I(Y;Z)$).} We utilised a Fano lower bound based on the classification error rate $e_T$ of the readout network on $K$-class data. Assuming the entropy of the representation is maximised given the error rate, the mutual information is lower-bounded by:
\begin{equation}
    I(Y;Z_T) \;\ge\; H(Y) - h_b(e_T) - e_T\log(K-1)
\end{equation}
where $H(Y)$ is the label entropy and $h_b(p)$ is the binary entropy function.

\textbf{Input Dependence ($I(X;Z)$).} For continuous latent codes, we estimated the superfluous information using a Gaussian proxy on whitened representations. Let $Z_T$ be the batch of latent codes at step $t$. We first whitened the representations to zero mean and unit variance. We then approximated the mutual information via the spectral properties of the code, specifically utilising the log-sum of variances under an isotropic noise assumption $\sigma^2$:
\begin{equation}
    \tilde{I}_\sigma(X;Z_T) = \frac{1}{2}\sum_{i=1}^{d} \log\left(1 + \frac{\lambda_i}{\sigma^2}\right)
\end{equation}
where $\lambda_i$ are the eigenvalues of the covariance matrix of $Z_T$. This metric serves as a trend indicator for the reduction of geometric volume occupied by the representations.

\subsection*{Cortical Refinement (Autoencoder) Model}
\textbf{Architecture.} We trained a convolutional autoencoder on various datasets such as the Fashion-MNIST ($28\times28$ grayscale). The encoder $g_\Phi$ consists of three convolutional layers ($32, 64, 64$ filters; $3\times3$ kernels; strides $2, 2, 1$) followed by a linear projection to a $d=32$ dimensional latent space. The decoder mirrors this structure. The readout network $f_\Omega$ is a two-layer MLP (hidden dimension 128) with LayerNorm, ReLU activation, and Dropout ($p=0.1$). The Refiner $\psi_\Psi$ is a 3-layer residual MLP (width 256) with layer normalisation and ReLU activations.

\textbf{Cross-Fit Protocol.} To prevent information leakage ($I(\Phi_S; S)$), the dataset was split into two disjoint sets: $\mathcal{D}_A$ (used to train the readout network $f_\Omega$) and $\mathcal{D}_B$ (used to train the refiner $\psi_\Psi$). The encoder $g_\Phi$ was pre-trained on $\mathcal{D}_A \cup \mathcal{D}_B$ and frozen.

\textbf{Refinement Objective.} The refiner updates the state as $z_{t+1} = z_t + \psi_\Psi(z_t)$. We consider two variants: a \textit{fixed} variant that always applies $T$ steps, and an \textit{adaptive} variant inspired by PonderNet~\cite{banino2021pondernet} that learns when to halt. In the adaptive variant, a learned halting head produces per-step stopping probabilities $p_t = \sigma(h(z_t))$, which define a categorical distribution over refinement steps: $q(t) = p_t \prod_{i<t}(1 - p_i)$, with residual mass assigned to step $T$. We trained the parameters $\Psi$ to minimise:
\begin{equation}
    \mathcal{L}_{\text{refine}} = \mathbb{E}_{(x,y)\in\mathcal{D}_B} \left[ \ell(f_\Omega(z_T), y) + \lambda_{\text{ctr}}\frac{1}{T}\sum_{t=0}^{T-1} \|z_{t+1}-z_t\|_2^2 + \beta \text{KL}(q(t) \| p_\lambda(t)) \right]
\end{equation}
where $\lambda_{\text{ctr}}$ penalises large latent jumps, and the third term (adaptive variant only) regularises the learned halting distribution $q(t)$ toward a truncated geometric prior $p_\lambda(t) = \lambda(1-\lambda)^{t-1}$ (normalised over $t = 1, \ldots, T$), encouraging efficient consolidation in fewer steps.

\textbf{Online Regularisation Baselines.} To rigorously test whether the benefits of consolidation were due to the mechanism of replay or simply increased single-pass regularisation, we trained ``Online-Only'' baselines with varying constraints. We replaced the deterministic bottleneck with a Variational Information Bottleneck (VIB) objective~\cite{alemi2016deep}, sweeping the regularisation strength $\beta$ (controlling $I(X;Z)$) across logarithmic and fine-grained scales ($\beta \in \{10^{-4}, \dots, 0.11, \dots, 4.0\}$). We also evaluated high-dropout variants ($p \in \{0.2, \dots, 0.8\}$). We defined the ``Fidelity-Generalisation frontier'' as the convex hull of test accuracy versus generalisation gap achieved by these single-pass models.

\subsection*{Predictive Coding Networks}
\textbf{Network Dynamics.} We implemented a hierarchical predictive coding network using an energy-based formulation (Langevin Predictive Coding~\cite{zahid2024langevin}). The model defines a variational free energy $\mathcal{F}$ consisting of a reconstruction term and a quadratic prior:
\begin{equation}
    \mathcal{F}(x, z) = \underbrace{\| \mathrm{Dec}_\Theta(z) - x \|^2}_{\mathclap{\text{Reconstruction Error}}} \; + \;\; \beta  \underbrace{\| z \|^2}_{\mathclap{\text{Prior Energy}}}
\end{equation}
where $\mathrm{Dec}_\Theta$ is the convolutional decoder network. Inference is performed via iterative gradient descent on the latent states $z$, with optional stochasticity (Langevin dynamics) to sample from the posterior:
\begin{equation}
    z_{t+1} = z_t - \eta \nabla_z \mathcal{F}(x, z_t) + \sigma \mathcal{N}(0, I)
\end{equation}
where $\eta$ is the step size and $\sigma$ is the injected noise scale. We used the same convolutional architecture as the Autoencoder experiments (3-layer ConvEncoder/Decoder) to ensure valid comparisons of representational capacity.

\textbf{Bidirectional Replay Protocol.}
To ensure a controlled comparison, we matched the computational budget ($T=10$ steps) and injected noise magnitude ($\sigma=0.05$) across both phases, varying only the information source and prior strength:
\begin{itemize}
    \item \textit{Wake (Perception):} Latents are initialised via a single forward pass through the encoder and refined for $T=10$ steps with weak priors ($\beta=0.001$). This simulates rapid, high-fidelity sensory encoding.
    \item \textit{Sleep (Consolidation):} A stored trace is projected top-down to generate a dream $\hat{x}$. This dream is re-encoded for $T=10$ steps with \textbf{increased prior weighting} ($\beta=0.1$) to simulate stronger homeostatic constraints and the dominance of internal priors over sensory precision.
\end{itemize}

\subsection*{Large Language Model Experiments}

\textbf{Architecture.} 
We utilised the Llama-3-8B-Instruct model~\cite{dubey2024llama} as the frozen backbone. The \textit{Cache Refiner} consists of lightweight Transformer blocks (hidden dimension 512) aligned to each layer of the backbone, as detailed in \cite{oomerjee:2025:bt}. The refiner interacts with the backbone's Key-Value (KV) cache via gated cross-attention. For an input sequence of length $L$, the backbone produces a cache $C \in \mathbb{R}^{L \times D}$. The refiner maps this to a compressed cache $C' \in \mathbb{R}^{L \times D}$ via in-place rewrites. Unlike standard compression methods that reduce dimensionality, our refiner maintains the vector space but optimises the information content.

\textbf{Consolidation Protocol.} 
Training was performed on the GSM8K and MATH datasets. To measure the generalisation benefit, we employed a two-phase protocol. First, the backbone was fine-tuned for 1 epoch (Phase 1). In Phase 2, the backbone was frozen, and only the Cache Refiner was trained for an additional epoch. We define the ``Generalisation Gap'' $\Delta_{Gen}$ as the difference between the validation loss (on novel templates) and the training loss (on seen templates):
\begin{equation}
    \Delta_{Gen} = \mathcal{L}_{val} - \mathcal{L}_{train}
\end{equation}
The refiner was trained to minimise the cross-entropy loss of the next reasoning step conditional on the rewritten cache. We observed that while $\mathcal{L}_{train}$ remained stable during Phase 2, $\mathcal{L}_{val}$ decreased significantly, effectively closing the gap.

\textbf{Key-Value Dissociation Analysis.}
To quantify representational drift, we computed the layer-wise cosine distance between the initial cache representations ($K_0, V_0$) and the consolidated representations ($K_t, V_t$) after $t$ refinement steps.
\begin{equation}
    d(x, y) = 1 - \frac{x \cdot y}{\|x\| \|y\|}
\end{equation}
We averaged this metric across all heads and tokens for Keys and Values independently. High distance indicates significant transformation (compression), while low distance indicates stability.

\textbf{Global Update Fraction.}
For each layer, we decomposed the refiner's update vectors into a shared component (the mean update across all tokens) and token-specific residuals. The Global Update Fraction is the proportion of total update energy attributable to the shared component.

\subsection*{Data Availability}
The datasets used in this study are publicly available. 
\textbf{OpenMathInstruct-2} is available at \url{https://huggingface.co/datasets/nvidia/OpenMathInstruct-2}. 
The mathematical reasoning benchmarks are available at their respective repositories: 
\textbf{GSM8K} (\url{https://github.com/openai/grade-school-math}), 
\textbf{MATH} (\url{https://github.com/hendrycks/math}), 
\textbf{SVAMP} (\url{https://github.com/arkilpatel/SVAMP}), 
\textbf{TheoremQA} (\url{https://github.com/TIGER-AI-Lab/TheoremQA}), 
\textbf{LogiQA} (\url{https://github.com/lgw863/LogiQA-dataset}), 
\textbf{Gaokao-MathQA} (\url{https://github.com/OpenLMLab/GAOKAO-Bench}), and 
\textbf{GSM-Hard} (\url{https://huggingface.co/datasets/reasoning-machines/gsm-hard}).

\subsection*{Code Availability}
Code to reproduce the experiments reported in this paper is available at
\url{https://github.com/zfountas/predictive-forgetting}.

\section*{Acknowledgements}
N.B. is funded by Wellcome (222457/Z/21/Z).

\bibliographystyle{unsrt}
\bibliography{references}

\newpage

\appendix

\setcounter{figure}{0} 
\renewcommand{\figurename}{Extended Data Fig.}

\section{Extended Data Figures}

\begin{figure}[h]
    \centering
    \includegraphics[width=0.75\linewidth]{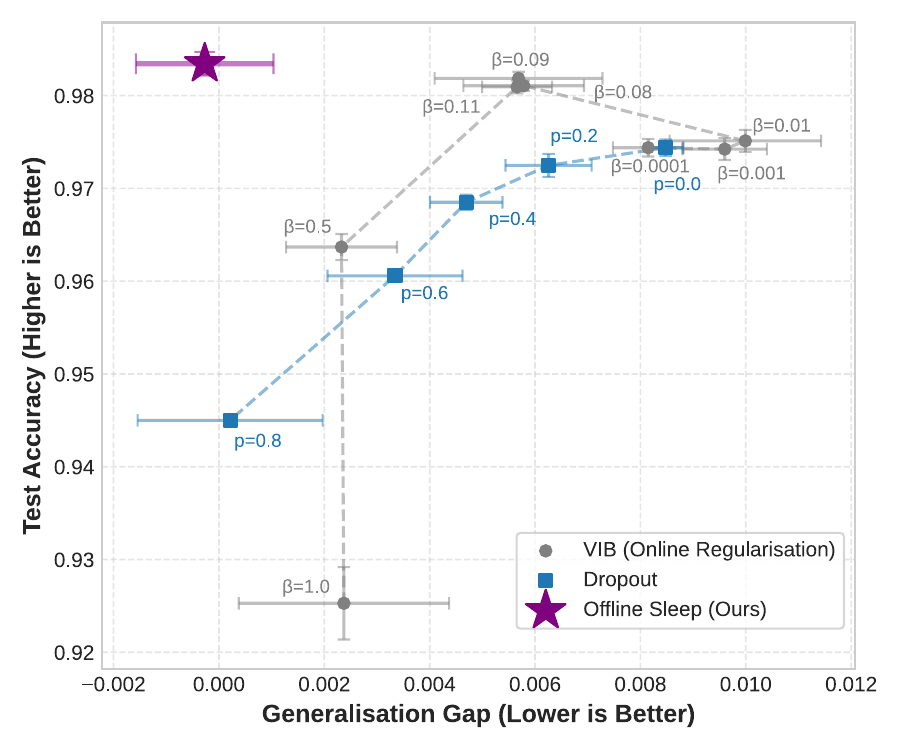}
    \caption{\textbf{The fidelity-generalisation frontier.} 
    To rigorously test whether the benefits of consolidation could be replicated by simply increasing regularisation during online training, we compared our Offline Replay model (Purple Star) against two single-pass baselines on the MNIST dataset. All models utilised the same convolutional architecture ($d=64$) described in Section~\ref{sec:ae-consolidation} to ensure a fair comparison of capacity.
    First, we trained an online agent with a Variational Information Bottleneck (VIB) objective (Gray circles), sweeping the regularisation strength $\beta \in \{10^{-4}, \dots, 10^0\}$ and performing a fine-grained sweep around the optimal region ($\beta \in \{0.08, \dots, 0.11\}$) to identify maximum performance.
    Second, we trained an online agent with dropout probabilities $p \in \{0.2, \dots, 0.8\}$ (Blue squares).
    The baselines define a convex ``fidelity-generalisation frontier'' (dashed lines), representing the unavoidable trade-off between minimising the gap and maintaining accuracy during single-pass learning. The Offline Sleep model lies beyond this frontier in the top-left quadrant. Data points represent mean $\pm$ standard deviation across $n=50$ independent seeds. This confirms that iterative offline refinement offers a computational advantage that cannot be recovered by online regularisation alone.}
    \label{fig:s1}
\end{figure}

\section{Supplementary Information: Refinement-with-Consolidation Experiments}

\subsection{Data and Splits}
We use MNIST, Fashion-MNIST (both $28{\times}28$ grey-scale, $K{=}10$), EMNIST-Balanced ($28{\times}28$ grey-scale, $K{=}47$), CIFAR-10, and SVHN (both $32{\times}32$ RGB, $K{=}10$).
Images are scaled to $[0,1]$ with \texttt{ToTensor}. 
From the official training split we hold out $10\%$ for validation. 
To prevent leakage from the refiner into the readout network's training statistics, we \emph{cross-fit} the original training set into two disjoint subsets: 
$\mathcal{A}$ (for training/evaluating the classifier readout and for reporting ``train'' metrics) and 
$\mathcal{B}$ (exclusively for training the refiner). 
Unless noted, the cross-fit ratio is $|\mathcal{B}|/|\mathcal{A}{\cup}\mathcal{B}|{=}0.5$ (abbreviated as \texttt{cf50}). 
We also report \texttt{cf25} ($0.25$) and \texttt{cf0} ($0$; disabled).

\subsection{Model Architecture}
\textbf{Encoder--Decoder (Phase 1).}
For grey-scale inputs the encoder is a 3-layer convolutional stack:
$\mathrm{Conv}(1{\rightarrow}32,3{\times}3,\text{stride}=2)$,
$\mathrm{Conv}(32{\rightarrow}64,3{\times}3,\text{stride}=2)$,
$\mathrm{Conv}(64{\rightarrow}64,3{\times}3,\text{stride}=1)$, each with ReLU.
The resulting $64{\times}7{\times}7$ feature map is flattened and mapped to a latent $z\in\mathbb{R}^{d}$ with a linear layer ($d{=}32$, unless stated).
For RGB inputs (CIFAR-10, SVHN) the encoder uses six convolutional layers ($3{\rightarrow}64{\rightarrow}64{\rightarrow}128{\rightarrow}128{\rightarrow}256{\rightarrow}256$, alternating stride $1$ and $2$) with batch normalisation after each layer.
The decoder mirrors the encoder via transposed convolutions (three for grey-scale, six for RGB) to reconstruct logits $\hat x$.
We train an \emph{autoencoder} with \textsc{bce-with-logits} reconstruction loss.%
\footnote{A VAE variant is supported but not used in the main sweeps; our analysis focuses on deterministic encodings used for the downstream readout network and refiner.}

\noindent\textbf{Classifier readout (Phase 2).}
A small MLP: $\mathrm{LayerNorm}(d)\rightarrow\mathrm{Linear}(d{\rightarrow}128)\rightarrow\mathrm{ReLU}\rightarrow\mathrm{Dropout}(p{=}0.1)\rightarrow\mathrm{Linear}(128{\rightarrow}K)$.

\noindent\textbf{Refiner (Phase 3).}
We study two refinement modes operating in latent space:
\begin{itemize}
  \item \textbf{Fixed} (\texttt{mode=fixed}): shared residual MLP block $f$ applied $T$ times:
  \[
    z_{t+1} \;=\; z_t \;+\; \tanh(\gamma)\, f(z_t),\quad t=0,\dots,T{-}1,
  \]
  where $f:\mathbb{R}^{d}\!\rightarrow\!\mathbb{R}^{d}$ is $\mathrm{LayerNorm}\!\rightarrow\!\mathrm{Linear}(d{\rightarrow}256)\!\rightarrow\!\mathrm{ReLU}\!\rightarrow\!\mathrm{Linear}(256{\rightarrow}d)$ and $\gamma$ is a learned scalar gate.
  The refined prediction uses $z_T$.
  \item \textbf{Ponder} (\texttt{mode=ponder}): same residual core as above plus a halting readout network $h$ producing per-step probabilities $p_t=\sigma(h(z_t))$ for $t{=}1..T$. 
  We form a truncated geometric posterior $q(t)$ over steps and predict with the mixture of class-probabilities across steps, weighted by $q(t)$.
\end{itemize}

\subsection{Training Procedure}
We use AdamW throughout, gradient clipping at $5.0$, early stopping on validation loss, and fixed seeds ($n{=}50$ per condition).

\paragraph{Phase 1 (autoencoder).}
optimise $\mathcal{L}_{\text{AE}}=\mathrm{BCEWithLogits}(\hat x,x)$ for $20$ epochs (lr $1\mathrm{e}{-3}$, wd $1\mathrm{e}{-4}$, patience $5$). 
We save the best checkpoint by validation loss and subsequently use the encoder’s \emph{deterministic} mapping $x\mapsto z_0$ for downstream phases.

\paragraph{Phase 2 (readout network).}
With the encoder frozen, optimise cross-entropy on $\mathcal{A}$ for $15$ epochs (lr $5\mathrm{e}{-4}$, wd $5\mathrm{e}{-4}$, patience $5$). 
We report train/test accuracy and generalisation gap $\mathrm{gap}=\mathrm{err}_{\text{test}}-\mathrm{err}_{\text{train}}$ at $t{=}0$ (no refinement).

\paragraph{Phase 3 (refiner).}
The refiner is trained \emph{only} on $\mathcal{B}$ (cross-fit) to avoid coupling the readout’s train metrics to the refiner updates. 
Loss is cross-entropy on the \emph{refined} prediction, with optional regularizers:
\[
\mathcal{L}_{\text{P3}} \;=\; \underbrace{\mathrm{CE}(\hat y^{\text{ref}},y)}_{\text{base}} 
\;+\; \beta_{\text{ponder}}\,\mathrm{KL}\big(q(\cdot)\,\|\,\pi_{\lambda}\big)
\;+\; \lambda_{\text{contract}}\cdot \textstyle\frac{1}{T}\sum_{t=0}^{T-1}\!\!\lVert z_{t+1}-z_t\rVert_2^{2},
\]
where the ponder KL (only in \texttt{mode=ponder}) penalizes deviation from a truncated geometric prior $\pi_\lambda$ with rate $\lambda\in(0,1)$, and the contractive penalty discourages large latent updates per step. 
We also optionally add small Gaussian noise to $z_0$ during training (\texttt{znoise}$=0.05$) to smooth refinements.
We use label smoothing $0.05$, lr $5\mathrm{e}{-4}$, wd $5\mathrm{e}{-4}$, patience $6$, $T\in\{1..4\}$ (plots sweep $T_{\text{eval}}{=}0..T$). 
Unless stated, the readout network remains frozen in Phase 3.

\subsection{Evaluation Protocol}
For each trained model we evaluate:
(i) the $t{=}0$ predictor (dashed curves) using $z_0$; 
(ii) the refined predictor (solid curves): for \texttt{fixed} we use $z_T$; for \texttt{ponder} we use the $q$-mixture of step-wise probabilities. 
We report train/val curves over Phase 3 epochs and conduct a sweep over the number of evaluation steps $T_{\text{eval}}\in\{0,\dots,T\}$ on the train and test splits. 
Following the cross-fit design, ``train'' metrics always refer to $\mathcal{A}$ (readout’s train set), never the refiner’s private $\mathcal{B}$.

\subsection{Mutual Information Estimators}
\paragraph{$I(Y;Z_T)$ (lower bound via Fano).}
Let $\mathrm{err}\in[0,1]$ be the classification error for a $K$-class task. 
Fano’s inequality gives $H(Y \mid Z_T)\le h(\mathrm{err})+\mathrm{err}\,\log(K-1)$, where $h$ is the binary entropy in nats.
Assuming a roughly balanced label marginal ($H(Y)\approx\log K$ for the balanced datasets used here), we bound
\[
  I(Y;Z_T) \;\ge\; \log K \;-\; \big[h(\mathrm{err})+\mathrm{err}\,\log(K-1)\big].
\]
We apply this to the measured errors from the $T_{\text{eval}}$ sweep on train and test, reporting mean$\pm$SEM across seeds.

\paragraph{$I(X;Z_T)$ (Gaussian proxy on whitened latents).}
To track how refinement alters the information that $Z_T$ can carry about $X$ without introducing a decoder-specific bias, we use a diagonal Gaussian proxy after per-dimension whitening:
if $Z_T$ (rows are samples) is whitened to zero-mean and diagonal variance $\{\sigma_i^2\}$ (empirical), and we posit an isotropic readout noise with std $\sigma$ at evaluation, we approximate
\[
  \tilde I_{\sigma}(X;Z_T) \;\approx\; \tfrac{1}{2}\sum_{i=1}^{d}\log\!\big(1+\sigma_i^2/\sigma^2\big).
\]
We keep a fixed whitening per seed (not per $T_{\text{eval}}$) to preserve comparability across steps. 
This yields a consistent \emph{proxy trend} versus $T_{\text{eval}}$; absolute values should be interpreted cautiously (diagonal and Gaussian assumptions). 
Where available, we compute this from saved representations $Z_T$ exported on train/val/test for each $T_{\text{eval}}$.

\subsection{Design Rationale}
\textbf{Cross-fit (\texttt{cf}).} 
By training the refiner on $\mathcal{B}$ while reporting ``train'' metrics on $\mathcal{A}$, we prevent the refiner from trivially improving the same examples used to measure training error, thereby stabilising estimates related to generalisation and controlling the effective dependence $I(\phi;\mathcal{A})$ of refiner parameters $\phi$ on the measurement set. 
We vary \texttt{cf} to show robustness.

\textbf{Contraction (\texttt{p3\_contract}).} 
Penalising step sizes $\|z_{t+1}-z_t\|_2^2$ encourages small, incremental refinements (reducing the risk of overfitting) and helps align the effective complexity of the refiner with the readout’s decision boundary.

\textbf{Ponder prior \& KL (\texttt{ponder\_lambda}, \texttt{ponder\_beta}).} 
In \texttt{mode=ponder}, the KL to a truncated geometric prior biases the halting distribution toward shorter computations unless evidence supports more steps, implicitly regularising computational depth.

\textbf{Label smoothing and $z$-noise.} 
Label smoothing ($0.05$) and light $z$-noise ($0.05$) further damp brittle refinements and reduce variance across seeds.

\subsection{Hyper-parameters and Ranges}
Unless overridden, we use the defaults in \ref{tab:hyper}.
We sweep $T\in\{1,2,3,4\}$, cross-fit ratio in \{0, 0.25, 0.5\}, and in some runs vary $\lambda_{\text{contract}}\in\{0,10^{-3}\}$ and \texttt{ponder\_lambda} $\in\{0.3,0.4,0.6\}$.

\begin{table}[h]
\centering
\caption{Key hyper-parameters (defaults).}
\label{tab:hyper}
\begin{tabular}{l l l}
\toprule
Component & Hyper-parameter & Default \\
\midrule
Data        & batch size & 128 \\
Encoder/Decoder (P1) & epochs / lr / wd / patience & 20 / $1\!\times\!10^{-3}$ / $1\!\times\!10^{-4}$ / 5 \\
Readout (P2)   & epochs / lr / wd / patience & 15 / $5\!\times\!10^{-4}$ / $5\!\times\!10^{-4}$ / 5 \\
Readout (P2)   & hidden / dropout & 128 / 0.1 \\
Refiner (P3, fixed/ponder) & $T$ (train) & 3 \\
Refiner (P3) & ref hidden & 256 \\
Refiner (P3) & epochs / lr / wd / patience & 20 / $5\!\times\!10^{-4}$ / $5\!\times\!10^{-4}$ / 6 \\
Refiner (P3) & label smoothing & 0.05 \\
Refiner (P3) & $z$-noise (train) & 0.05 \\
Refiner (P3) & contract $\lambda_{\text{contract}}$ & 0.0 (optionally $10^{-3}$) \\
Ponder mode  & $\lambda$ (geometric prior) & 0.3 \\
Ponder mode  & $\beta_{\text{ponder}}$ (KL weight) & 0.01 \\
Cross-fit    & $|\mathcal{B}|/|\mathcal{A}{\cup}\mathcal{B}|$ & 0.5 (cf50) \\
Latent       & $d$ & 32 \\
MI proxy     & $\sigma$ (for $\tilde I_\sigma$) & 0.1 \\
\bottomrule
\end{tabular}
\end{table}

\subsection{Reported Quantities}
From Phase 2 we report $t{=}0$ train/test accuracy and generalisation gap. 
From Phase 3 we report: (i) train/val curves (both $t{=}0$ and refined), (ii) test accuracy vs.\ $T_{\text{eval}}$, (iii) test gap vs.\ $T_{\text{eval}}$, (iv) $I(Y;Z_T)$ lower bound vs.\ $T_{\text{eval}}$, and (v) $\tilde I_\sigma(X;Z_T)$ vs.\ $T_{\text{eval}}$ where representations are available. 
All curves show mean$\pm$SEM across seeds.

\subsection{Reproducibility Artifacts}
For each run we save: \texttt{args.json}, \texttt{phase1\_history.json}, \texttt{phase2\_summary.json}, \texttt{phase3\_history.json}, \texttt{phase3\_summary.json}, and \texttt{phase3\_Tsweep.csv}. 
Optional latent dumps for MI are stored as \texttt{repr/<split>/Teval\_\{k\}/zT.npy}. 
Plots and aggregated CSVs are produced by our analysis scripts (multi-seed grids, $T_{\text{eval}}$ sweeps, and MI summaries).

\end{document}